\documentclass[preprint,authoryear,twocolumn]{elsarticle}

\usepackage[utf8]{inputenc}
\usepackage{setspace}%% The amsthm package provides extended theorem environments
\usepackage[utf8]{inputenc}
\usepackage[english]{babel}
\usepackage[T1]{fontenc}
\usepackage{graphicx}
\usepackage{latexsym}
\usepackage{textcomp}
\usepackage{amsfonts}
\usepackage{amsmath}
\usepackage{color}
\usepackage{dcolumn}
\usepackage{bm}
\usepackage{amssymb}
\usepackage{epstopdf}
\usepackage{xcolor}
\usepackage{enumerate}
\usepackage{multirow}
\usepackage{wrapfig}
\usepackage[round]{natbib}
\usepackage{float}
\usepackage{lineno,hyperref}
\usepackage{subcaption}
\usepackage[flushleft]{threeparttable}

\definecolor{myred}{rgb}{0.8,0,0}
\definecolor{myblue}{rgb}{0,0.3,0.6}
\definecolor{mygreen}{rgb}{0,0.4,0}

\begin{document}

%\nocite{TitlesOn}
\begin{frontmatter}

\onecolumn
\title{Quantification of the morphological characteristics of hESC colonies}

\author[1,*]{Sirio Orozco-Fuentes}
\author[2,3]{Irina Neganova}
\author[1]{Laura E. Wadkin}
\author[1]{Andrew W. Baggaley}
\author[4]{Rafael A. Barrio}
\author[2]{Majlinda Lako}
\author[1]{Anvar Shukurov}
\author[1]{Nicholas G. Parker}

\address[1]{School of Mathematics, Statistics and Physics, Newcastle University, NE1 7RU, United Kingdom.}
\address[2]{Institute of Genetic Medicine, Newcastle University, NE1 7RU, United Kingdom.}
\address[3]{Institute of Cytology, Russian Academy of Sciences, St-Petersburg, Russia.}
\address[4]{Instituto de Física, Universidad Nacional Autónoma de México, Mexico City, Mexico.}
\address[*]{Corresponding author: sirio.orozco-fuentes@newcastle.ac.uk}

%\keywords{Keyword1, Keyword2, Keyword3}
% % \linenumbers
% \setcounter{secnumdepth}{1}
\begin{abstract}

The maintenance of the pluripotent state in human embryonic stem cells (hESCs) is critical for further application in regenerative medicine, drug testing and studies of fundamental biology. Currently, the selection of the best quality cells and colonies for propagation is typically performed by eye, in terms of the displayed morphological features, such as prominent/abundant nucleoli and a colony with a tightly packed appearance and a well-defined edge. Using image analysis and computational tools, we precisely quantify these properties using phase-contrast images of hESC colonies of different sizes (0.1 -- 1.1$\, \text{mm}^2$) during days 2, 3 and 4 after plating. Our analyses reveal noticeable differences in their structure influenced directly by the colony area $A$. Large colonies ($A > 0.6 \, \text{mm}^2$) have cells with smaller nuclei and a short intercellular distance when compared with small colonies ($A < 0.2 \, \text{mm}^2$). The gaps between the cells, which are present in small and medium sized colonies with $A \le 0.6 \, \text{mm}^2$, disappear in large colonies ($A > 0.6 \, \text{mm}^2$) due to the proliferation of the cells in the bulk. This increases the colony density and the number of nearest neighbours.

We also detect the self-organisation of cells in the colonies where newly divided (smallest) cells cluster together in patches, separated from larger cells at the final stages of the cell cycle. This might influence directly cell-to-cell interactions and the community effects within the colonies since the segregation induced by size differences allows the interchange of neighbours as the cells proliferate and the colony grows. Our findings are relevant to efforts to determine the quality of hESC colonies and establish colony characteristics database.

\end{abstract}

\begin{keyword}

Human embryonic stem cells \sep Cellular characteristics \sep Feature extraction \sep Voronoi diagram \sep Segregation

\end{keyword}
\end{frontmatter}

% \linenumbers
% \doublespacing
% \thispagestyle{empty}
\section{Introduction}

\label{intro}

Human embryonic stem cells (hESCs) are pluripotent cells, derived from the blastocyst-stage embryos, which have the capacity to differentiate and give rise to all tissues of the body. More than 20 years ago, a method to derive stem cells from human embryos was discovered and allowed the use of these cells for further research in vitro \citep{ThomsonScience}. They provide an opportunity to study early human development and the processes by which pluripotency is lost and differentiation into different tissues occurs \citep{Pera5515}. The specific signalling factors promoting stem cells to remain unspecialised in culture without differentiation have been highly optimised during the last two decades \citep{VazinFreed}. Also, protocols have been developed to differentiate hESCs towards all three germ layers for disease modelling, cell-based therapies and drug screening \citep{AnastasiaBoehm}. After the derivation of human induced pluripotent stem cells (hiPSCs) \citep{TAKAHASHI2007861}, which made creating patient-matched embryonic stem cell lines feasible, hESCs and hiPSCs have become an emerging model for developmental studies and personalised medicine \citep{SOLDNER2018615,Odorico2004,Zhu2013}. 

Although the genetic and signalling pathways that control pluripotency in hESCs have been described in the last decade \citep{NeganovaOnco,NeganovaCDK1, Dalton2013,NeganovaCDK2,ZHAO2017141,doi:10.1002/stem.2954}, much less is known about the factors that control the arrangement of the cells into a pluripotent colony and how this affects pluripotency. Human ESCs grow as a multicellular colony. At the single cell level the transcription factors (TFs) associated with the maintenance of pluripotency fluctuate stochastically \citep{BUGANIM20121209,LiBelmonte}. These different expression states are maintained by different signalling, transcriptional, and epigenetic regulatory networks. However, pluripotency, considered as an emergent property of stem cell populations and their niches (rather than a property of single cells), is controlled by niche-mediated regulation in response to mechanical, chemical and physical stimuli \citep{Peerani4744,ViningMooney}. Thus, understanding how pluripotency is affected by cell segregation within the bulk of a colony is of practical importance in generating and selecting the optimum clones, and automating this for industrial-scale production.

The current paradigm in stem cell biology dictates that pluripotency is regulated by a set of TFs, such as NANOG, OCT4 and SOX2, that fluctuate stochastically along the cell cycle and show a heterogeneous expression at the single cell and colony levels \citep{Torres-Padilla2173,Stahlberg2162,WolffInheritance,MESSMER2019815}. A quantitative analysis of the arrangement of the cells within the colonies is a prerequisite for the construction of hypothesis-driven mathematical and computational models that can provide explanations for the observed dynamics in hESC colonies and their regulation at the microenvironment level.

To characterise how hESCs regulate their assembly into a multicellular colony we performed a detailed quantitative analysis of hESC colonies of different sizes during the exponential growth phase, at days 2, 3 and 4 after plating. These quantitative properties of the colony morphologies are poorly studied, while the pluripotent regulation at the expression level captures most of the attention \citep{warmflash2014, ETOC2016302, Nemashkalo3042, WolffInheritance}.

Previous works within our group have demonstrated that isolated hESCs growing on Matrigel\texttrademark \ with mTESR1 media are highly motile ($\sim 16.25 \, \mu$m/h) and sensitive to the presence of nearby cells \citep{Wadkin1,Wadkin2}. As these single cells proliferate and possibly aggregate with other cells, colonies are formed. Within a colony, the cells collectively move in the bulk, and the whole structure remodels due to cell division and cell spreading. If the conditions for the maintenance of pluripotency are lost, the cells differentiate \citep{10.1371/journal.pone.0006082,NeganovaOnco}.

The collective motion of cells within colonies involves mechanical and biochemical interactions between the cells. The most important physical factors affecting the cell dynamics are the proliferation and short-range interactions with other cells, driven by the pressure flow due to the mitotic expansion. Research on self-renewal and differentiation of hESCs indicates that both processes are highly influenced by biophysical signals, such as the mechanical forces and the rigidity of the extracellular matrix (ECM) \citep{SunFu}.
\cite{OsafunehESCdensity} demonstrated the role of local cell density and colony size, showing an increase in the correlation of differentiation efficiency of both hESC and hiPSC cell lines towards pancreatic cells. This indicates that the local density attained within the cultures should be highly relevant for the maintenance of pluripotency.  Moreover, a recent study for human induced pluripotent stem cells (hiPSCs) has shown the migratory behaviours of the cells vary on different substrates (e.g., laminin, fibronectin, matrigel) due to changes in their adhesion properties, concluding that the regulation of the motility of the cells might improve the clonality of the forming colonies \citep{LI20102442,CHANG201927}. With the colony growth, the cells regulate each other through cell-cell and cell-media interactions \citep{Peerani4744,XuDing}  resulting in community effects that regulate pluripotency and differentiation \citep{Nemashkalo3042}. Therefore, the results obtained at the single-cell level cannot be sufficient to deduce biological processes at the colony level as a whole.

The formation of hESC colonies in vitro is a natural process emerging when a single cell proliferates and forms a small cluster. In this work, we analyse colonies growing in feeder-free conditions (Matrigel\texttrademark \ ). It has been demonstrated that cells within the same colony have a higher correlation of being of the same type, e.g. pluripotent or primed towards differentiation \citep{Nemashkalo3042}, which might be achieved by the combination of endogenous signals between the cells and extrinsic factors (addition of differentiation cues).
 % , where single individuals entwine their lamellipodial protrusions enabling them to stick closely together.

The size and morphology of the colonies provide with preliminary information about the pluripotency status of the cells. The undifferentiated state is assessed through the specific morphology of the cells and the colonies, see Table \ref{table1}, which is typically estimated visually. The morphological features of pluripotent hESCs inside a colony are: roundness, large nucleus, scant cytoplasm and prominent (highly visible) nucleoli. As the colony grows, the central part becomes more compact than the periphery. Human ESCs colonies with areas $A < 0.3 \, \text{mm}^2$ show white spaces or gaps between the cells \citep{atlashPSCs2, atlashPSCs}. To account for these features, we measured the individual cells by manually outlining their nucleus. This gives us the overall nuclei morphology in terms of the colony size and day after plating. Several properties, such as nucleus area and shape descriptors (aspect ratio, Feret's diameter, circularity, roundness and solidity), defined in the supplementary material (SM)) were measured. Our results show that colonies with $A < 0.1 \, \text{mm}^2$ show distinctive features in their structural properties, such as a large nucleus cell area and a large separation between nearest neighbours. Both quantities decrease as the colony size increases, with the largest colony showing the smallest value in the mean cell nucleus area. This may be caused by the pressure build-up due to mitotic events in the bulk of the colony.  To measure the segregation of the small (recently divided) cells, we introduce a segregation order parameter. Our results suggest the self-organisation of the cells in terms of their nucleus sizes, since the small cells cluster together in patches, separating the larger cells from each other. Recent results by \cite{Nemashkalo3042}, using micropatterned colonies of a few cells, indicate that interactions between neighbours can lead to sustained and homogeneous signalling for differentiation. Taking these two results together, a continuous interchange of neighbours between the cells as they grow and divide, might affect their cellular communication and ultimately the pluripotent state attained by the colony as a whole. 

% Colonies that show well-defined nucleoli and nuclei were selected

\begin{table}
    \begin{center}
\caption{Morphological features of hESCs and their colonies.}
\label{table1}
\begin{tabular}{ |c|c| }
% % \label{tab:widgets}
\hline 
{\bf Type}  &{\bf Characteristics} \\ \hline\hline
\multirow{4}{*}{Cell}   & Prominent nucleoli   \\
  & Scant cytoplasm \\
  & Round \\
  & Small   \\ \hline
\multirow{4}{*}{Colony} & Round \\
  & Flat  \\
  & Well-defined edges \\
  & Gaps between the cells ($A < 0.6 \, \text{mm}^2$)\\ \hline
\end{tabular}
    \end{center}
\end{table}

%{\color{red} To address: why hESCs grow as a colony? What is known about this?}

\begin{figure*}
\begin{center}
    \includegraphics[width=0.95\textwidth]{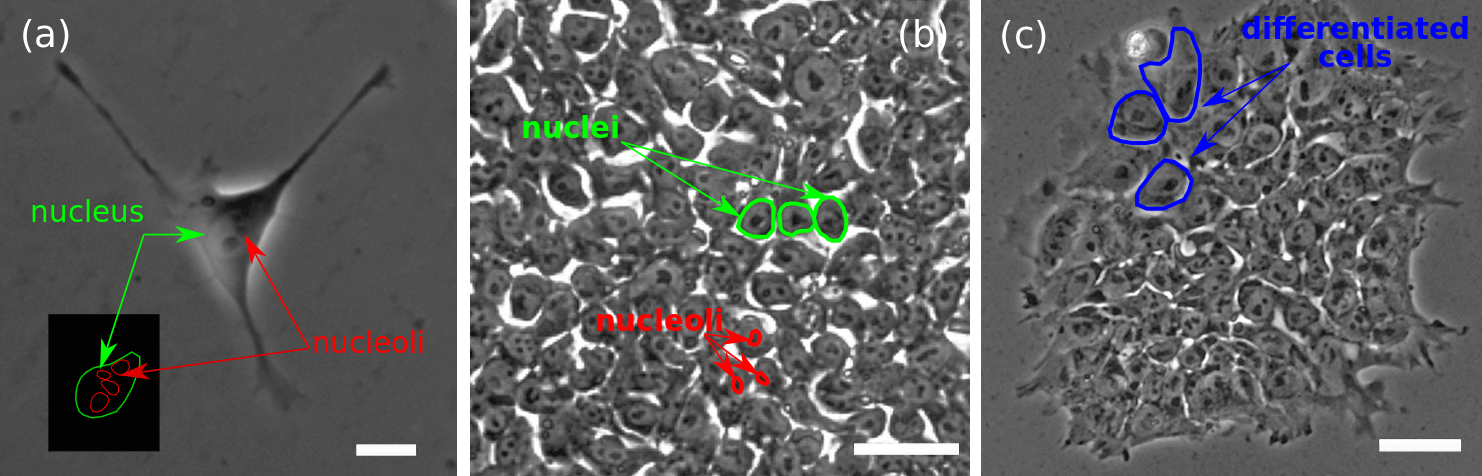}
\end{center}
\caption{(a) Phase-contrast image of a single isolated hESC at day 2 after plating, showing a well-defined nucleus, nucleoli (black dots) and spreading lamellipodia. Bar $20 \, \mu$m. (b) Detail of the spatial arrangement of cells within a colony, with well-defined nuclei and prominent nucleoli (black dots). The very distinctive gaps between the cells occur in colonies with areas $A  < 0.6 \, \text{mm}^2$. Bar $50 \, \mu$m. (c) HESC colony in which the cells located at the top-left (outlined in blue) are probably differentiated since they show a larger nuclei. Note how the differentiated cells have a different nucleus/cytoplasm ratio compared to the undifferentiated cells. Bar $100 \, \mu$m.}
\label{nucleus_outline}        
\end{figure*}

Computational models are helpful to quantitatively analyse and improve the understanding of the processes that underlie fate decisions in hESCs and hiPSCs. However, before establishing the appropiate protocol for {\it in silico} approaches, it is important to quantify the morphological features frequently used in the visual identification of pluripotent hESC colonies, see Table \ref{table1}, in agreement with previous publications  \citep{johkura,atlashPSCs,Ullmann} . These give us value information about the morphological properties of the cells arranged in colonies. In the future, this information will be integrated alongside other mechanisms that determine the behaviour of the system, to build algorithms of interaction rules aiming to understand their emergent properties \citep{Herberg2250}.

% , which will possibly facilitate the identification of the regulatory principles controlling self-renewal and differentiation in hESCs and hiPSCs.

% These parameteres can then be used to analy

 % will assist in the development of a mathematical tool to standardise these morphological parameters. This tool can then be used to analyse the pluripotency status of colonies, as a non-invasive method to choose the best clone or colony.

% The structure of the paper is as follows, in  \S\ref{expImaging} we propose a non-invasive image-based evaluation method to characterise hESCs within colonies, the Voronoi diagram. In \S\ref{results} we show the results 

\section{Materials and Methods}

%Capitalize trade names and give manufacturers' full names and addresses (city and state). 

\label{expImaging}
\subsection{Cell culture and propagation}

Human embryonic stem cells (hESCs) (H9 cell line, WiCell, Madison, WI) were passaged on 6-well plates coated with hESC-qualified Matrix at a 1:4 split ratio using an EDTA-based dissociation solution. 2 ml of mTERSR1 media was used per wel. The cells were kept in small clumps avoiding the passaging of single cells (due to low rates of survival). We aimed to plate cell aggregates of approximately 15 -- 20 cells each. The culture was kept for 4 days at $37 \, ^\circ$C with a humidified $5 \% \,\text{CO}_2$ atmosphere. The colonies were imaged at day 2, 3 and 4 after plating before they reached a $60 \, \%$ confluency across the well. 

% During the sub-culturing (passaging) of hESCs,  the cells were transferred from a previous well into a new well with a fresh growth medium, enabling the propagation of the cell line.

The ability of hESCs cells to form colonies depends on the cytoskeleton rearrangement, contraction of actin filaments, the interaction between the cells, and the timely function of regulatory proteins. When isolated, the cells have their cytoskeleton and lamellipodia unfolded and spreading over the substrate, see Figure \ref{nucleus_outline}(a). In colonies, the cells are close to each other as shown in Figure \ref{nucleus_outline}(b). This section of a colony contains several cells in which the nuclei, nucleoli (dark spots) and gaps (white spaces between the cells) are easily detected. Larger and denser colonies do not show gaps and the cells are closer to each other, see Figure \ref{scheme_outline}.

% For any cell type in culture, there is a characteristic growth pattern that follows three distinct well-defined phases. The first one, termed the lag phase, starts just after plating and is a period of slow growth when the cells are adapting to their environment and preparing to grow. During the second phase, known as the logarithmic phase, the cells proliferate exponentially and consume the nutrients in the growth medium and 
% rapidly fill the available substrate. Afterwards, a stationary phase follows in which there is a growth limiting factor, such as the lack of nutrients. Cultures of hESCs are usually sub-cultured before the second phase ends.

% HESCs after plating, attach to the substrate which provides adhesion and contact dependent factors necessary to maintain undifferentiated cell type. The culture medium provides nutrients and growth factors that enable the cells to grow in flat dense colonies with a steady proliferation. 

We studied the colonies using phase contrast microscopy, since this method allows the cells to behave as naturally as possible without the need to stain the cells with fluorescent dyes which may induce photo-toxicity \citep{IchaPhotoxicity} and possible changes in cell behaviour \citep{LulevichDyes}. During the last day of image acquisition, the confluency of the cells was about $60 \, \%$, meaning that most of the colonies did not merge with each other and were quasi-bidimensional structures.

% We selected colonies of vary size with morphological features typical of  undifferentiated colonies, i.e., with clear borders, containing small round cells with large nuclei and notable nucleoli \citep{atlashPSCs}. The internal structure of a single isolated cell is shown in Figure \ref{nucleus_outline}(a) with a scheme at the bottom-left side outlining the nucleus and nucleoli. These cells are tightly packed within colonies, see Figure \ref{nucleus_outline}(b) and show a well-defined nucleus and nucleoli.  The Figure \ref{nucleus_outline}(c) shows a hESC colony in which the cells located at the top-left (outlined in blue) are differentiated. The cells show a change in their nucleus/cytoplasm ratio.

Examples images of hESC and their colonies are shown in Figure \ref{nucleus_outline}.  The internal structure of a single isolated cell is shown in Figure \ref{nucleus_outline}(a) with a scheme at the bottom-left side outlining the nucleus and nucleoli. Colonies of varying size were selected with morphological features typical of undifferentiated colonies, i.e., with clear borders, containing small round cells with large nuclei and notable nucleoli \citep{atlashPSCs}; an example is shown in Figure \ref{nucleus_outline}(b). Figure \ref{nucleus_outline}(c) shows a hESC colony with cells showing a different nucleus/cytoplasm ratio.

\subsection{Imaging of hESCs}
\begin{figure*}
    \begin{center}
        \includegraphics[width=0.98\textwidth]{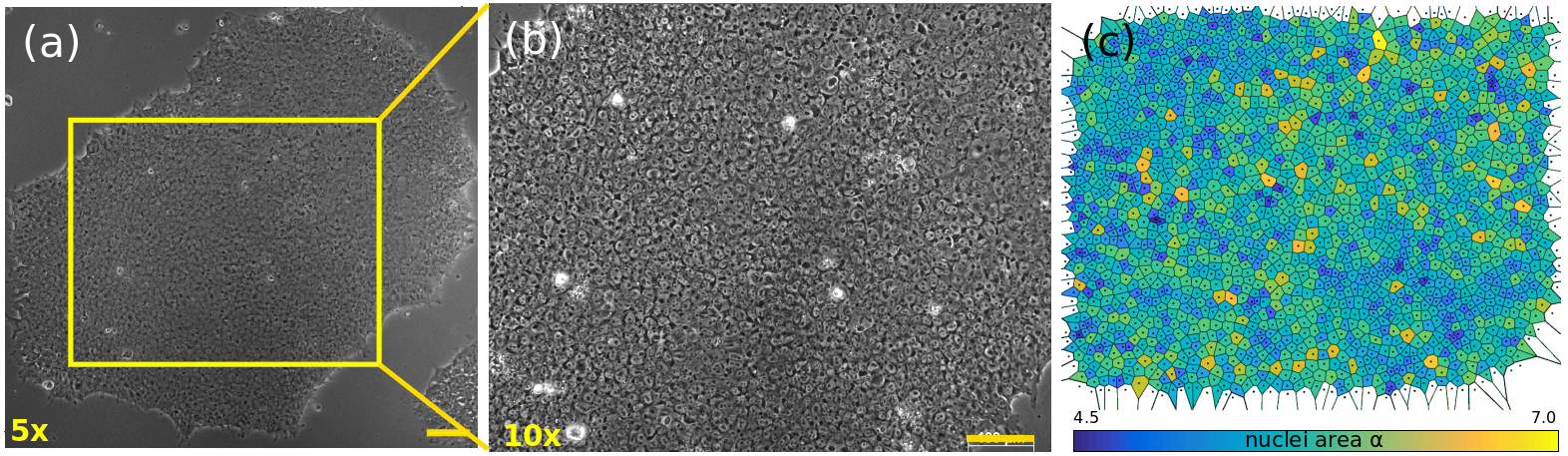}
        \caption{Phase-contrast images of a hESC colony at day 3 after plating. For large colonies we used (a) low magnification (5$\times$) to capture the boundaries and (b) higher magnification (10$\times$) to measure the cells features (the enclosed region in (a)). This colony is a densely packed example with no gaps within the cells. Bars $100 \, \mu$m. (c) Voronoi tessellation obtained from the centroid position of the cells. The nuclei area is shown in logarithmic scale.}
        \label{scheme_outline}   
    \end{center}
\end{figure*}
%analysed with an area $A = 1.132 \, \text{mm}^2$

To quantify the morphological characteristics within the colony we outlined each nucleus manually and extracted several parameters such as the centroid position, nucleus area and relevant shape descriptors included in ImageJ \citep{Imagej}. During mitosis, the cells adopt a spherical shape, detach from the ECM, divide and reattach again, with the two new daughter cells lying in close proximity to each other. We recorded these mitotic events during the manual tracing of the cell nuclei. 

% In large colonies, the cells are tightly packed and for some cases, it was difficult to have a clear visualisation of the individual cells while ignoring other cell structures such as the lamellipodia; under those circumstances, the cells were ignored. 

% Our goal is to quantify the morphological features frequently used in laboratory during visual inspection to identify pluripotent colonies of hESCs and summarized in Table  and standardise with a mathematical tool the morphological parameters of hESCs colony used to analyse their pluripotent status and generate with this tool parametric morphological characteristics, as non-invasive methods to choose the best clone or colony.

We outlined the nuclei of the cells in $19$ colonies of different sizes (see Table \ref{tableres} in the \hyperref[SuppMaterial]{SM} for further details). Alongside this information, the boundaries of 38 colonies were obtained using an edge detection algorithm through a canny Deriche filtering \citep{Imagej}, see Table \ref{tableres2} in the \hyperref[SuppMaterial]{SM} for more details. An example of the analysis performed on the colonies is shown in Figure \ref{scheme_outline}(a). This sample has an area  $A = 1.132 \, \text{mm}^2$ and it was imaged at day 3 after plating. For large colonies, we imaged the structure at low magnification (5$\times$) to account for the colony's features, and at a higher magnification (10$\times$) focusing in the bulk, Figure \ref{scheme_outline}(b), to outline the cell nuclei, Figure \ref{scheme_outline}(c). Using ImageJ \citep{Imagej} software (http://
rsb.info.nih.gov/ij/), we processed the outlined images and obtained the centroid position, area, perimeter and shape descriptors (aspect ratio, solidity, circularity and Feret's diameter) of each nucleus and, at a larger scale, of each colony.

\subsection{Voronoi diagram}
\label{sec:VD}

% Our spatial data analyses are based on the Voronoi diagram (VD) that subdivides the embedding space into polyhedral cells or regions. The construction of the VD of a colony, shown with an example in Figure \ref{voroscheme}, starts with a set of seed nodes: the nuclei positions. The space between these seeds is then partitioned in such a way that each point in the space is assigned to its closest seed. 

The spatial data analyses presented in this work are based on the Voronoi diagram (VD) that divides the area in the most equalitarian fashion, in such a way that the area occupied by a cell is obtained by tracing straight lines between the position of a cell and all its neighbours and drawing a perpendicular line in the middle. These perpendicular lines form a convex polyhedron, called the Voronoi cell. Therefore the VD is the collection of Voronoi cells. The generated "cells" are not uniform in shape and their number of faces vary from one to another.

The geometric dual of the VD is called the Delaunay triangulation (DT). It connects those points of a VD that share a common border. The VD facilitates spatial analysis, e.g., the closest neighbours identification through the adjacency matrix, and is used in many fields of science, including cell biology, \citep{Honda78,Saito82}. We used the VD to measure the structural properties of the colonies and the DT to obtain the intercellular distances.

\begin{figure}
  \begin{center}
    \includegraphics[width=0.5\textwidth]{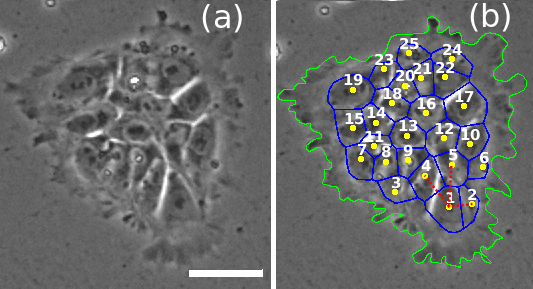}
  \caption{(a) The VD for a small colony with 25 cells and (b) constructed through the set of centroid positions of the cells. The dotted (red) lines show the first nearest neighbours for cell 1. The green line is the outline of the colony border. Scale bar: $50 \, \mu$m.}
      \label{voroscheme}   
    \end{center}
\end{figure}

As an example, Figure \ref{voroscheme} shows the VD for a small colony with 25 cells. The nearest neighbours of cell 1 are connected with dotted (red) lines (DT), i.e., cells 2, 4 and 5, giving the distance to the nearest neighbours. The cells 3 and 6 are the second nearest neighbours of cell 1. We performed this analysis on larger colonies, see Figure \ref{scheme_outline}(a). An example of the Voronoi tessellation obtained for the region in  Figure \ref{scheme_outline}(b) is shown in Figure \ref{scheme_outline}(c) for 1982 cells. The nuclei are coloured according to the logarithm of the nucleus area ($\alpha$) to ease in the visualisation (see colorbar). Using the centroids as input points, we obtained the VD, shown with black continuous lines. Since the cells are round in shape (see \hyperref[SuppMaterial]{SM}, Figure \ref{ShapeDescriptorsNuclei}), the VD allows accurate identification of each cell's neighbourhood. Through the DT, we identified the nearest neighbours associated to each cell and calculated the mean number of nearest neighbours $\langle N_n \rangle$\footnote{From now on, the angular brackets $\langle \rangle$ will denote the average taken over the cell population within a given colony. The bar $-$ will denote the average taken over several colonies.} and the mean distance to nearest neighbours (or intracellular distance) $\langle \ell_n \rangle$, measured from the centroid position of the two cell pairs.

\section{Results}
\label{results}

% \subsection*{Morphological analysis}

After plating, hESCs form small clusters of several cells attached to each other and to the ECM. These are the initial seeds from which larger colonies start to grow through proliferation. The cells inside the colonies display self-propulsion, resulting in a movement of the colony as a whole through the culture. We assume events in which cells from other colonies travel through the ECM and attach to other colonies are rare for hESCs plated at low densities. Depending on the initial plating density, merging of colonies might occur after some time. In our experiments, the confluency of the colonies was less than 60\% on day 4. Due to the variability in single cell movements, cell growth and mitotic events, the biophysical interactions between the cells in the colonies are not distributed uniformly. As a result, the colonies become more irregularly shaped when the number of cells increases.

\subsection{Nucleus morphology}
\label{NucleiMorph}

\begin{figure}
    \begin{center}
        \includegraphics[width=0.4\textwidth]{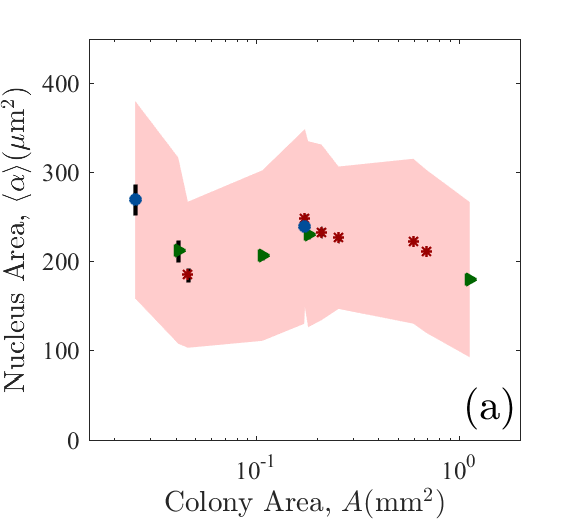}
        \includegraphics[width=0.4\textwidth]{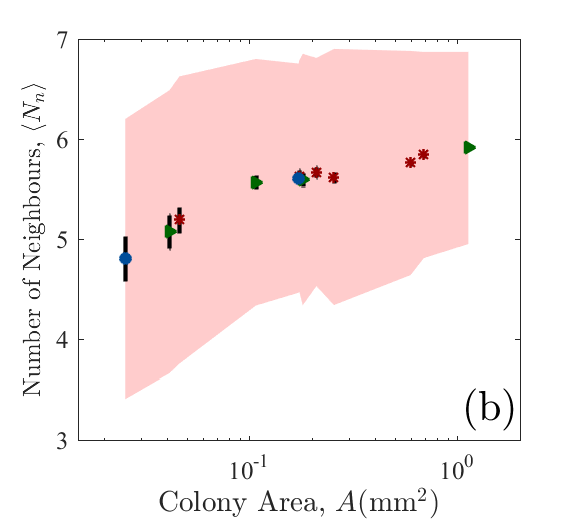}
        \includegraphics[width=0.4\textwidth]{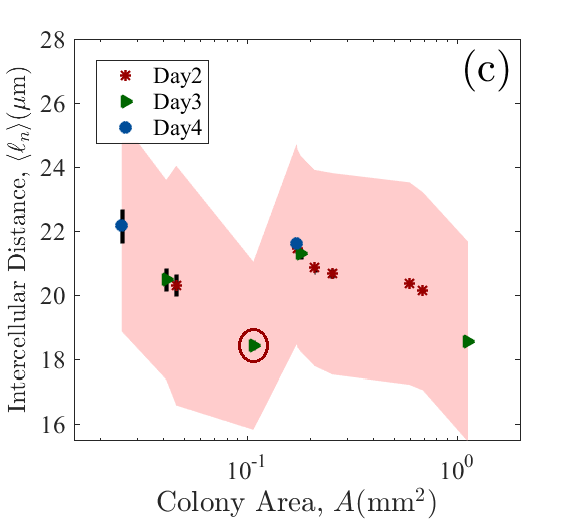}
    \end{center}
    \caption{Mean values of the (a) nucleus area $\langle \alpha \rangle$, (b) the number of nearest neighbours $\langle N_n \rangle$, and (c) the intercellular distance $\langle \ell_n \rangle$ as a function of the colony area $A$. The standard deviation range is the shaded region. The standard error of the mean is shown as a black line around the data point. Data points are presented with different symbols according to the day at which the image was taken (see legend in panel c).}
\label{plotsTable}      
\end{figure}

The mean nucleus area $\langle \alpha \rangle$, mean number of nearest neighbours $\langle N_n \rangle$ and mean intercellular distance $\langle \ell_n \rangle$ of the colonies are shown in Figure \ref{plotsTable} as a function of colony area $A$. The mean cell nucleus area $\langle \alpha \rangle$, Figure \ref{plotsTable}(a), shows high variability between colonies of different sizes and sampling days. The smallest colonies, with $\sim 70$ cells at day 3 ({\color{mygreen} $\blacktriangleright$}) and 25 -- 46 cells at day 4 ({\color{myblue} $\bullet$}), have the largest mean nucleus area, with $\langle \alpha \rangle = 269 \pm 111 \, \mu$m$^2$ and $212 \pm 104 \, \mu$m$^2$ respectively. The small colonies analysed at day 2 with $\sim 115$ cells ({\color{myred} $\ast$}) result in $\langle \alpha \rangle = 184 \pm 82 \, \mu$m$^2$, a lower value than their later imaged counterparts. Although our error bars are large since we included cells undergoing mitosis, this result indicates that the cell nucleus is larger in the small colonies formed at later stages of passage, {\it e.g} day 4 colonies with a few dozen cells.

The mean number of nearest neighbours $\langle N_n \rangle = 4.8 \pm 1.4$ cells (day 4), 5.1 $\pm$ 1.4 cells (day 3) and 5.2 $\pm$ 1.4 cells (day 2) decreases with $\langle \alpha \rangle$ and increases with the colony area $A$, whilst the mean intracellular distance $\langle \ell_N \rangle$, decreases with $\langle \alpha \rangle$ as expected. For small colonies at day 4 showing large cell nucleus areas, our results indicate that cell-to-cell contacts, occur on average with less than five cells that have a larger separation between them. This may influence the regulation of community effects in colonies of these sizes.

% A single colony analysed at day 3 with $N_c = 305$ and $A = 0.107 \, \text{mm}^2$, with an average nucleus area $\langle \alpha \rangle = 206 \pm 96 \, \mu \text{m}$ and number of nearest neighbours $\langle N_n \rangle = 5.57 \pm 1.23$  shows a peculiarly short intracellular distance $\langle \ell_n \rangle = 18.43 \pm 2.61 \, \mu \text{m}$, see the datapoint outlined with a red circle in Figure \ref{plotsTable}(c). The presence of floating cells on top of the colony as shown in Figure \ref{illColony} might indicate anomalies in this sample.

\begin{figure}
    \begin{center}
        \includegraphics[width=0.4\textwidth]{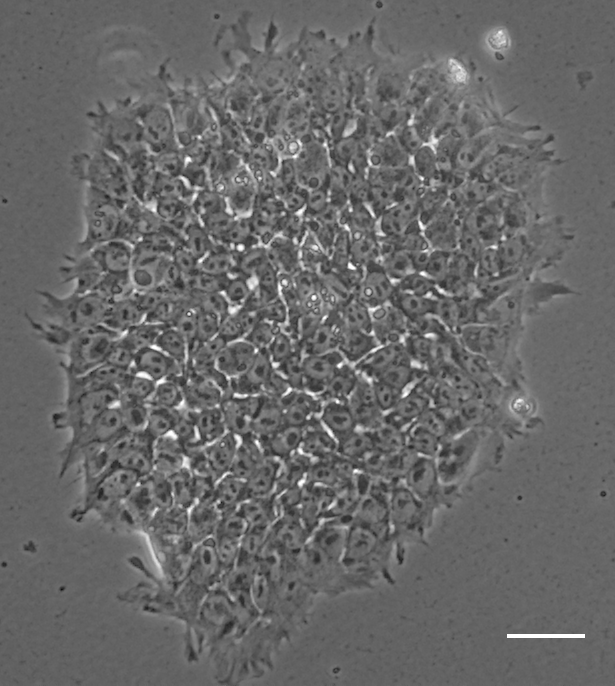}
    \end{center}
    \caption{Colony with an area $A= 0.107 \, \text{mm}^2$ and $N_c = 305$ cells imaged at day 3. The mean nucleus area is $\langle \alpha \rangle = 206 \pm 96 \, \mu \text{m}^2$ and the mean intercellular distance is $\langle \ell_n \rangle  = 18.43 \pm 2.61 \, \mu \text{m}$. Bar $50 \, \mu \text{m}$.}
\label{illColony}      
\end{figure}

The day 3 colony highlighted with the red circle in Figure \ref{plotsTable}(c) has a particularly short intracellular distance for its area.  The image of this colony, shown in Figure \ref{illColony}, shows several cells floating on top of the colony; this might indicate the interplay of other factors in the rearrangement of the cells in this sample.

    % \caption{Colony with an area $A= 0.107 \, \text{mm}^2$ and $N_c = 305$ cells imaged at day 3. The mean nucleus area is $\langle \alpha \rangle = 206 \pm 96 \, \mu \text{m}^2$ and the mean intercellular distance is $\langle \ell_n \rangle  = 18.43 \pm 2.61 \, \mu \text{m}$. The presence of floating cells on top of the colony might indicate anomalies in this sample, such as cells undergoing senescence or apoptosis. Bar $50 \, \mu \text{m}$.}
%unfavourable growth conditions

For colonies containing between 350 and 550 cells, we obtain $\overline {\langle \alpha \rangle} = 239 \pm 104 \, \mu$m$^2$. The nucleus area is almost constant and shows only modest variability between days 2 and 3. The mean number of neighbours $\langle N_n \rangle$ increases steadily with the colony area, Figure \ref{plotsTable}(b), whilst the mean intercellular distance $\langle \ell_n \rangle$ decreases, i.e. the colony becomes more compact, Figure \ref{plotsTable}(c).

Finally, for the largest colonies analysed (the last three points to the right in Figures \ref{plotsTable}), there is a clear decrease in the mean nucleus area with colony size. The largest colony has the highest mean number of neighbours and the smallest inter-cell distances. Therefore, for large colonies (higher density) on average six neighbours are involved in cell-to-cell interactions. Visually these colonies are very dense, cells are tightly packed and there are no gaps between them. 

In summary, colonies with $A < 0.1 \, \text{mm}^2$ and $N_c < 100$ cells have the largest nuclei and intercellular distances with less neighbouring cells. On the other hand, the largest colony has cells with the smallest nuclei, short intercellular distances and six neighbours on average. Since the number of nearest neighbours increases with the colony area (and the number of cells) we suggest that the number of cells in a colony increases faster than its area as the cells fill the space within the colony.

We measured several other properties for the nuclei, such as their aspect ratio, perimeter, Feret's diameter, circularity, roundness and solidity, as shown in Figure \ref{ShapeDescriptorsNuclei} and \ref{ShapeDesc} in the \hyperref[SuppMaterial]{SM}. Some of these parameters have been used to characterise mouse embryonic stem cell (mESCs) colonies during differentiation \citep{mescsFeret}. However, for hESCs, at the single and colony level, these measurements do not show any significant change in behaviour that would indicate changes in the the morphology of the cells and colonies in terms of the days after plating and colony sizes, see Supplement Tables \ref{shapedescriptors} and \ref{tableres2}.

\subsection{Probability distribution functions of nuclei area}

\begin{figure}
    \includegraphics[width=0.42\textwidth]{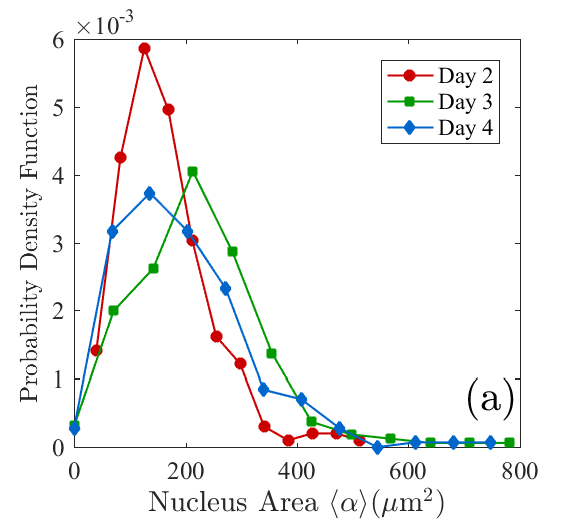}
     \includegraphics[width=0.42\textwidth]{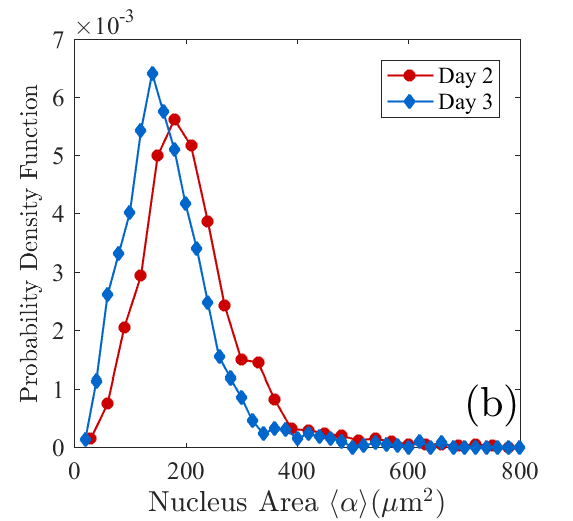}
\caption{Probability density function (PDF) for the nucleus area ($\alpha$) measured for (a) colonies imaged at day 2, 3 and 4, with an area $A < 0.1 \, \text{mm}^2$ and (b) the largest colonies imaged at day 2 and 3.}
\label{pdfAreas}      
\end{figure}

The size and shape of the cells are good indicators of their health and most importantly of their viability as a pluripotent cell for stem cell research. The averages of the quantities obtained in the previous section give a rough estimation of the behaviour of these variables in terms of the colony sizes. However, to account for the variability of the nuclei areas within a colony, we calculated the probability distribution functions PDF of $\alpha$, shown in Figure \ref{pdfAreas}, for several samples, dividing them according to sampling day and size.

Colonies with $A < 0.2 \, \text{mm}^2$ (day 4) show an abrupt change in $\langle \alpha \rangle$ as a function of sampling day. Both day 3 and 4 colonies have a broader distribution, with cells having nuclei of sizes $\alpha > 600 \, \mu$m$^2$. The colonies at day 2 have a narrower distribution in which all nuclei have an area $\alpha < 500 \, \mu$m$^2$. It is important to keep in mind that day 3 and 4 colonies have half as many cells as the day 2 colonies, see Table \ref{tableres} in the \hyperref[SuppMaterial]{SM} for further details. Overall, we observed a similar PDF for the smallest colonies, which becomes narrower as the cells increase their numbers. Therefore, a large nucleus in small samples may be due to a lack of compactness and pressure between the cells.

For the largest colonies analysed at days 2 and 3, with $\langle N_c \rangle$ = 1257 $\pm$ 327 and 1982 cells, respectively, the distributions become narrower as the colonies get bigger, see Figure \ref{pdfAreas}(b). Occurrences of nuclei area $\sim 350 \, \mu \mbox{m}^2$ disappear, and overall the cells become more homogeneous in size with $\langle \alpha \rangle = 179 \pm 87 \, \mu \mbox{m}^2$ for the day 3 colonies, see Figure \ref{scheme_outline}. Under these circumstances, crowding effects due to mechanical cell competitions may take place in the bulk. %The internal pressure in the bulk of the colony, generated by the local cell proliferation may become exceedingly large, which can lead to changes in the cell cycle and after a critical pressure cell cycle arrest.

% The effect of mechanical forces on stem cell differentiation is dependent upon cell type as well the phenotype/environment it is in. For example, mechanical compression significantly increases the chondrocytic expression of bone-marrow-derived MSCs embedded in a hydrogel; however, embryonic stem cell-derived embryoid bodies significantly downregulate chondrocytic gene expression under the same conditions (Terraciano et al., 2007). Taken together it is clear that mechanical forces, at least in part, regulate stem cell differentiation with the differential effects dependent upon a number of cell specific factors.

% \caption{Probability density function (PDF) for the nucleus area ($\alpha$) measured for (a) colonies imaged at day 2, 3 and 4, with an area $A < 0.1 \, \text{mm}^2$ and $\langle N_c \rangle$ = 39 $\pm$ 10, 72 $\pm$ 3.5 and 115 $\pm$ 1.4 cells, respectively and (b) the largest colonies imaged at day 2 and 3, with $A = 0.643 \pm 0.067$ and $1.131 \, \text{mm}^2$  ($\langle N_c \rangle$ = 1257 $\pm$ 327 and 1982 cells). The maximum of the PDF displaces to the left as the colonies get bigger, which indicates that the cells become smaller possibly due to pressure build-up inside the colonies due to cell proliferation.}

\subsection{Colony morphology}

During colony formation, there are physical forces transmitted through the cells that affect the local mechanical properties and, therefore, play important roles in cellular behaviour such as adhesion properties, cell proliferation, differentiation and death (through the activation of biochemical signals) \citep{Taylor,Ehrlicher,Fletcher,HanLiu}. The colony shape is one of the qualitative features used to identify the best colonies and best clones. To quantify their form, we obtained the area $A$, perimeter $P$ and shape descriptors of 38 colonies, see Supplement Table \ref{tableres2} in the \hyperref[SuppMaterial]{SM}. To measure changes in cell and colony morphologies as the cell numbers increase, we counted the cells in 19 colonies and added these results to the other 19 colonies analysed in the previous section. 

\begin{figure}
\begin{center}
    \includegraphics[width=0.4\textwidth]{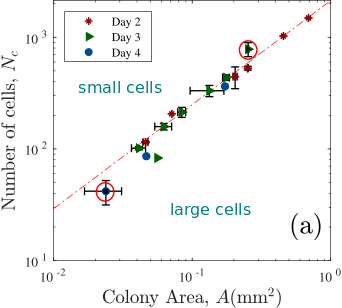}
    \includegraphics[width=0.4\textwidth]{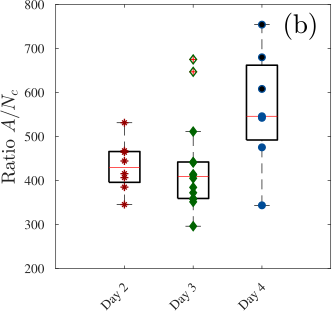}
\end{center}
\caption{(a) The number of cells $N_c$ as a function of the colony area $A$ (log-log scale). Data points are coloured according to the number of days after plating in which the image was taken. The red dotted line corresponds to the best fit to $N_C(A) = \kappa A^\beta$, with a scaling factor $\kappa = 2130$ and exponent $\beta = 0.93$ ($\text{R}^2$ = 0.97). The three outliers for $A < 0.01 \, \text{mm}^2$ correspond to colonies with distinctive features when compared to the rest (partly differentiated). (b) Mean area per single cell, $ \overline{A} = A/N_c$ are the following $\langle A/N_c \rangle = 433 \pm 57\, \mu$m$^2$ (day 2), $434 \pm 109 \, \mu$m$^2$ (day 3) and $564 \pm 135 \, \mu$m$^2$ (day 4). The medians are shown as red central lines on each box, which edges represent the 25th and 75th percentiles respectively. Some data points at day 3 (crossed on top with {\color{red} +}) were considered outliers. The points that correspond to colonies with differentiated cells are filled in black.}
\label{AreavsNCP}
\end{figure}

Figure \ref{AreavsNCP}(a) shows the the number of cells $N_c$ as a function of  colony area $A$. 
A power function trend line, $N_c = \kappa A^\beta$, is appropriate with scaling factor $\kappa=2130$ and exponent $\beta = 0.93$ ($\text{R}^2$ = 0.97), see red dotted line. The exponent $\beta$ is approximately one, which corresponds to the cells maintaining the same nucleus area while the colony grows. The two largest colonies at day 2, with $A = 0.456 \, \text{mm}^2$ and $0.691 \, \text{mm}^2$, follow the same trend. However, small colonies from day 4 (left red circle) and some samples on day 3 (right red circle) deviate from this relationship. 

% Figure \ref{AreavsNCP}(b),  shows the space available for each cell in each colony measured as a function of the day. The box on top of the samples represents the mean (central red line), and the 25th and 75th percentiles (box edges). The data points filled in black at day 3 are considered as outliers by the program (Matlab 2018b\textregistered) which applies a formal test assuming the data points follow a normal distribution. These data points correspond to the samples shown in Figure \ref{AreavsNCP}(a) inside the right red circle. 

% boxplot draws points as outliers if they are greater than q3 + w × (q3 – q1) or less than q1 – w × (q3 – q1), where w is the maximum whisker length, and q1 and q3 are the 25th and 75th percentiles of the sample data, respectively.

% The default value for 'Whisker' corresponds to approximately +/–2.7σ and 99.3 percent coverage if the data are normally distributed. The plotted whisker extends to the adjacent value, which is the most extreme data value that is not an outlier.

%jellou

\begin{figure*}
    \begin{center}
        \includegraphics[width=0.98\textwidth]{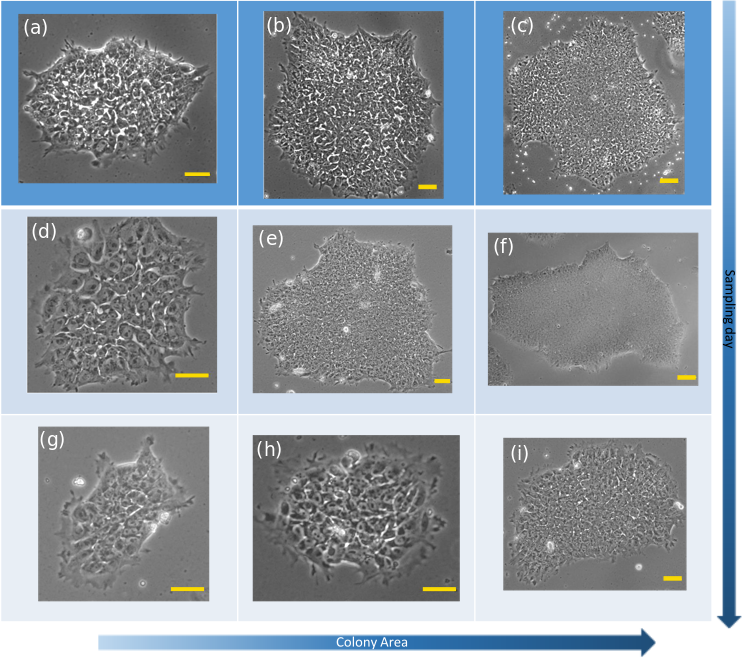}
        \caption{(a-c) Phase-contrast images of hESC colonies analysed at day 2 with areas $A = 0.071, \, 0.254$ and $0.691 \text{mm}^2$, respectively. (e-f) Day 3 colonies with areas  $A = 0.056, \, 0.255$ and $0.555 \text{mm}^2$ and (g-i) day 4 colonies with  $A = 0.025, \, 0.047$ and $0.173 \text{mm}^2$ Bars $50 \, \mu$m in a, b, d, e, g, h, i and  $100 \, \mu$m in c and f.}
        \label{colonies-analysed}   
    \end{center}
\end{figure*}

We detect five small colonies, imaged at day 4 (with $N_c < 100$), whose overall behaviour indicated exceedingly large nuclei, see Figures \ref{nucleus_outline}(c) and  \ref{colonies-analysed}(d) in the \hyperref[SuppMaterial]{SM}. Those colonies are highlighted in Figure \ref{AreavsNCP}(b) with black points and are detected as outliers by the boxplot method. A more detailed analysis of these colonies indicates that a proportion of the cell population has undergone differentiation, see Figure \ref{nucleus_outline}(c). 

% At day 4 we obtained colonies with slightly larger cells, which may be an indication that changes in the cell sizes have occurred. 

% The mean area available per cell $\overline{A} = A /N_c$ as a function of the day of plating, considering the outliers at day 3 are the following: $\langle \overline{A} \rangle_{\text{day 2}} = 432.7 \pm 57.3$, $\langle \overline{A} \rangle_{\text{day 3}} = 433.7 \pm 109.1$ and $\langle \overline{A} \rangle_{\text{day 4}} = 564.4 \pm 134.7$. The data points filled in black at day 4 correspond to the colonies showed in Figure \ref{AreavsNCP}(a) inside the left red circle. 

Comparing colonies of similar sizes measured at day 2, $A = 0.252 \pm 0.002 \, \text{mm}^2$, and day 3, $A = 0.254 \pm 0.003 \, \text{mm}^2$, we observe that the former has fewer cells ($N_c$ = $528 \pm 20$) than the latter ($N_c = 782 \pm 112$). Therefore, there is an increment of  $\Delta N_c \approx 250$ cells in the bulk of the colony without an increase in the colony area. We can infer that the disappearance of the gaps between the cells, highly visible at day 2, is a result of newly dividing cells filling the voids. Consequently, the power-law relationship (linear on log-log scale) shown in Figure \ref{AreavsNCP}(a) between $A$ and $N_c$ holds only for colonies at day 2 and some colonies on day 3.

%boxplot draws points as outliers if they are greater than q3 + w × (q3 – q1) or less than q1 – w × (q3 – q1), where w is the maximum whisker length, and q1 and q3 are the 25th and 75th percentiles of the sample data, respectively.

%% GAP DISAPPEARANCE

Colonies with $A < 0.3 \, \text{mm}^2$ show gaps between the cells. For $A > 0.3 \, \text{mm}^2$ these gaps start to disappear in the middle of the colony and are completely lost for colonies with areas $A > 0.6 \, \text{mm}^2$. It is known that hiPSC colonies form actin (a linear polymeric micro-filament) fences encircling the colony that exerts extensive mechanical stress to enforce colony morphology and compaction \citep{narva}. We suppose that at initial stages of colony formation the cells accommodate themselves in such way that they have a higher intercellular distance between them, without being tightly packed, forming a polymeric fence around them to enforce compactness. For small and medium-sized colonies with gaps, there should be an outward pressure flow of cells at the boundary in order to accommodate newly divided cells in the bulk while keeping these spaces empty.

With the increase in cell numbers, we assume that there are more mitotic events in the colony and less time to re-organise the colony edges. Therefore it is possible that the fences formed at previous stages continue to maintain a strong adhesion at the border with the ECM, making the filling of gaps possible. 
% For large colonies, there should be crowding effects that could lead to cell cycle arrest.

\subsection{Segregation and population mixing}

%Segregation is an ubiquitous phenomenon observed in a myriad of systems at different length scales, from granular matter to cells aggregated in tissues.  

Segregation of cells in tissues and during pattern formation is an important phenomenon that occurs during the early phase of embryonic development, which ends with the formation of the three germ layers \citep{Kurosaka}. The arrangement of cells in the embryo occurs due to changes in the environment (surface cues) that induce differences in adhesion properties and changes in the cytoskeleton \citep{Fagotto3303}. These differences in adhesion properties between neighbouring cells maintain a physical separation between different cell types, and it is one of the basic mechanisms for the pattern formation during development and wound healing \citep{KRENS2011189}.  Although {\it in vivo}, migration of hESCs is responsible for the segregation into physically distinct regions after a few rounds of divisions, {\it in vitro}, the presence of migratory effects is undesirable due to population mixing and loss of clonality \citep{CHANG201927}. 

From experiments on isolated hESCs, our measurements indicate that the cell grows until it reaches a size of $\sim 300 \, \mu$m$^2$ (unpublished results), after which it divides into two almost identical cells of sizes $150 \, \mu$m$^2$. Although hESCs form quasi-flat colonies, they show a highly dynamic behaviour in the bulk, with cells constantly migrating, interchanging neighbours and displacing each other due to mitosis. Specifically, it is at this stage that the cells detach from the culture, divide and re-attach again, resulting in an important factor that promotes cell mixing within the population which possibly affects the level of pluripotency achieved within the colony. To account for segregation in hESC colonies, we explore if the newly divided (small) cells are driven away from larger cells in the colony, which may be caused by the pushing of larger cells.

To measure if the small cells are segregated from the largest cells in the bulk of the colony, we introduce a segregation order parameter depending on the level of separation between small (type A) and large (type B) cells. Several order parameters can be introduced to characterise processes of segregation according to several segregation criteria \citep{Rivas2011}.

The VD, see Section \S\ref{sec:VD}, identifies accurately the number of nearest neighbours in each colony.  We introduce a suitable segregation order parameter that depends explicitly on the number of nearest neighbours. We consider two types of particles A and B, if the system is segregated, each particle A will have more neighbours of the same type. The segregation order parameter $\delta$ is defined as follows,

\begin{equation}
    \delta = 1 - \frac{N_c N_\text{AB}}{N_n N_\text{A} N_\text{B}},
    \label{segregationOP}
\end{equation}

\noindent
where $N_\text{AB}$ is the sum of the number of A Delaunay neighbours that B particles have, double counting the A particles that are neighbours of different B particles and $N_n$ is the number of nearest neighbours that each particle has on average. For a perfectly mixed system, with $N_n = 6$ Delaunay neighbours, equation (\ref{segregationOP}) results in $\delta \approx 0$. If the system is completely segregated, for example, one cluster of A particles surrounded by other of B particles, $\delta \sim 1$. The calculation of $\delta$ was performed for the largest colonies analysed each day for which the mean number of nearest neighbours exceeded $\langle N_n \rangle > 5 $, see Figure \ref{plotsTable}(b).% i.e. making the use of equation (\ref{segregationOP}) possible. 

\begin{figure*}
\begin{center}
    \includegraphics[width=0.98\textwidth]{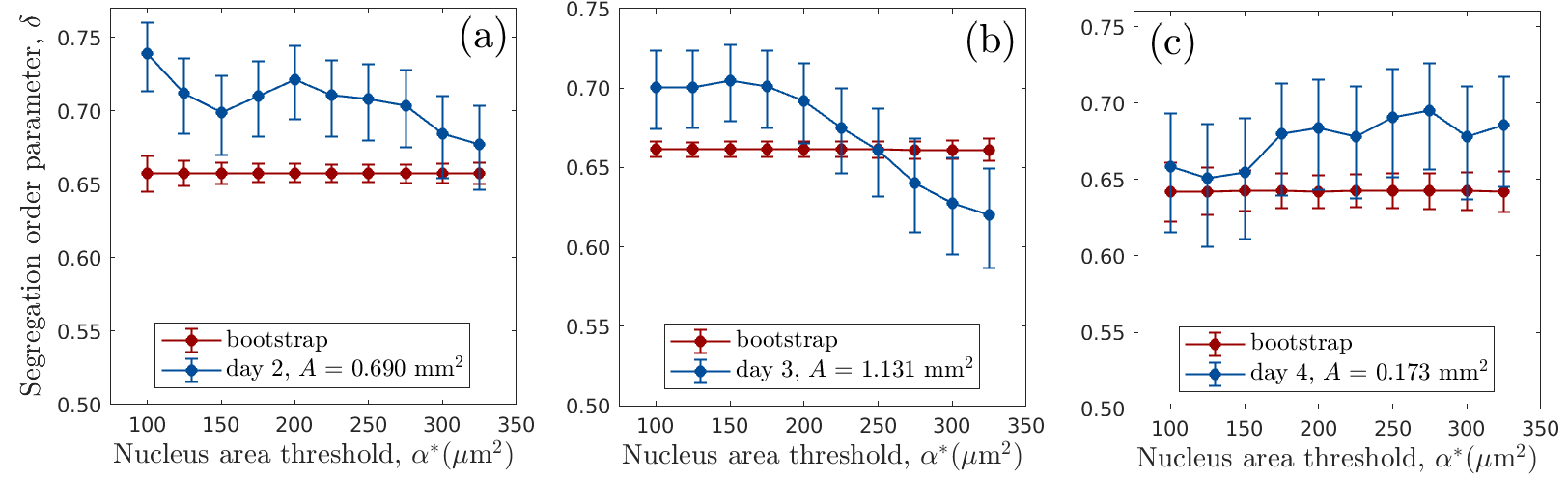}
  \end{center}
\caption{Segregation order parameter $\delta$ for the colonies with areas (a) $A= 0.690 \, \text{mm}^2$ (day 2), (b) $A = 1.131 \, \text{mm}^2$  (day 3) and (c) $A= 0.173 \, \text{mm}^2$ (day 4). The segregation is calculated in terms of an area threshold $\alpha^\ast$ as a proxy for two cell types. Type A cells ($\alpha < \alpha^\ast = 200 \, \mu$m$^2$) are segregated from the larger type B cells in (a) and (b). The results are inconclusive for c. The values of $\delta$ obtained by re-sampling the data sets (bootstrap method) are shown alongside these results, see the legend in the inset.}
\label{segregationCol}  
\end{figure*}

We use an area threshold $\alpha^\ast$ to group the cells into two categories: type $A$ cells ($\alpha < \alpha^\ast$) and type B cells ($\alpha \ge \alpha^\ast$) and vary $\alpha^\ast$ between $100 \, \mu$m$^2$ and $325 \, \mu$m$^2$. Applying equation (\ref{segregationOP}) to the largest colonies measured each day, gives the results shown in Figure \ref{segregationCol}. We also show the results of the bootstrap to detect differences with a re-sampled data set. The colonies with areas $0.690 \, \text{mm}^2$ (day 2) and $1.131 \, \text{mm}^2$ (day 3) contain $N_c  = 1489$ and $1982$ cells respectively, panel a and b, whereas the colony with area $0.173 \, \text{mm}^2$ have $N_c=363$ cells, panel c. Our results strongly suggest that, in large colonies, small cells  ($\alpha < 200 \, \mathrm{\mu m}^2$) are segregated from larger ones. In both cases, see panels a and b in Figure \ref{segregationCol}, the curve for $\delta$ is above the one obtained from the bootstrap method.  For $\alpha^\ast > 275 \, \mu$m$^2$, in Figure \ref{segregationCol}(b), $\delta$ reaches lower values that the bootstrap; this means that cells with $\alpha > 275 \, \mu$m$^2$ have less chance of having neighbours of the same size, and therefore, they are surrounded by smaller particles which are clustered together. As we increase $\alpha^\ast$ beyond $200 \, \mu \mbox{m}^2$, $\delta$ decreases reaching the values for the random configuration. Therefore, for the largest colony, Figure \ref{segregationCol}(b), the cells are separated in a random fashion at $\alpha^\ast$ = $250 \, \mu \mbox{m}^2$. However, for $\alpha^\ast$ > $300 \, \mu \mbox{m}^2$, $\delta$ continues decreasing until it reaches $0.66$, which indicates that the larger cells within the colony (mitotic events) tend occur far apart from each other. The results shown in Figure \ref{segregationCol}(c), for the colony with $A = 0.173 \, \text{mm}^2$ remain within the values obtained with the bootstrap method and the results are inconclusive, suggesting the need of larger colonies to obtain accurate measurements.

% The Figure \ref{segregationCol}(d) is calculated for a small colony with $A = 0.173 \, \text{mm}^2$ (day 4) and $N_c = 363$ cells. We also show the results of the bootstrap to detect differences with a re-sampled data set. Our results for the largest colony, see panel c, strongly suggest that cells with $\alpha < 200 \, \mu$m$^2$, are segregated from larger cells, since the curve for $\delta$ is above the curve obtained from the bootstrap method.

 % The colonies with areas $A= 0.596 \, \text{mm}^2$ (day 2), $0.690 \, \text{mm}^2$ (day 2), $1.131 \, \text{mm}^2$ and $0.173 \, \text{mm}^2$ contain more than $N_c > 1000$.

% For $\alpha^\ast = 200 \, \mu$m$^2$, t

\section{Discussion}

We quantified the morphological and structural properties as well as the behaviour of hESCs during colony formation. Human embryonic stem cells self-organise into colonies with sharp edges and a strong adhesion at the border that promotes the maintenance of the pluripotent state by keeping the colonies tightly packed \citep{GINIS2004360,narva,doi:10.1002/stem.2954}. Our analyses reveal that the colonies change their morphological properties as the cells proliferate and the whole colony becomes larger.

Colonies with $A < 0.3 \, \text{mm}^2$  show white spaces between the cells that disappear, starting from the middle of the colony and finally at the edges, as the colony grows. There is a decrease in the nuclei size with an increase of colony size, accompanied by an increase in the number of first nearest neighbours available for each cell. Each cell within the largest colony, interacts, by contact, on average with six other cells. We suggest that this may have implications in the establishment of pluripotency on the local environment.

After these gaps are completely filled, we observe in large colonies the emergence of collective effects driven by the constant pushing and pulling of cells which drive the smallest and newly divided individuals to cluster in patches within the colony. %cellular quiescence. 
Since the analysed colonies were grown on Matrigel\textsuperscript{TM}, their migratory effects are large \citep{Wadkin1,Wadkin2} and this could be a relevant factor in the spatial organisation of the cells within the colony. The continuous re-organisation of the colonies implies that neighbours are interchanged continually and consequently the cell population is continuously mixed; this directly influences the level of clonality within the colonies and the outcome of community effects that will furthermore influence the pluripotency achieved by the population \citep{CHANG201927, Nemashkalo3042}. 

Recent studies on hiPSCs with modified molecular regulators of cortical tension and cell-cell adhesion (through target genes ROCK1 and CDH1, respectively) have shown the emergence of distinct patterning events within hiPSC colonies through cell-driven segregation that dictated the colony organisation without the loss of pluripotency \citep{elife2}. Our results indicate that newly divided (small) cells are driven away from larger cells, clumped together in patches. Whether this effect is solely due to the mechanical effects (pushing) between the cells or changes in the cells' cortical tension/adhesion properties along the cell cycle remains unknown and its elucidation requires further work.

\section{Conclusion}

The morphological analysis of hESC colonies is a powerful non-invasive tool to evaluate their quality and choose the best clones for medical applications, unlike invasive labelling procedures that involve genetic manipulation. Although the implementation of an algorithm for the automatic detection of cells within a colony was beyond the scope of this work, once such a method is developed, the parameters estimated throughout this paper can be easily implemented at a larger scale, to quantify accurately the parametric properties of pluripotent colonies.

Our work indicates that the mean nuclei area and mean distance between nearest neighbours might be good parameters to detect changes in the morphology of the colonies, despite the inherent variability in the cell sizes associated to the cell growth and the cell cycle. Our algorithms detect that small colonies at day 4 show distinctively larger cell nuclei and intercellular distances. These changes in their morphological properties might affect their pluripotency levels. Assuming an average hESC cycle duration of 14.6 h \citep{WolffInheritance} and an exponential colony growth starting from a single founder cell, we estimate that day 4 colonies with $N = [25, 46]$ cells, were formed between 2--3 days before imaging. Therefore, later formed colonies have cells with changed morphological characteristics. Following this same premise, the larger colonies analysed at day 2 and 4 with thousands of cells, most certainly did not started from single founder cells. Our results suggests that this might be advantageous for the maintenance of their structural properties.
The segregation of the cells inside the colony has strong biological implications in regards of the genetic and phenotypic spreading, since neighbouring individuals eventually end up in completely different locations.
% $N_c = N_c^0 \exp^{\kappa t}$,
 % we obtain a growth factor $\kappa = 0.05$, $N_0 = 1$ and 

% So, what is conclusion from this work based on parameters you measured? 
% If on day 3 you can found colony with about 40 cells -is that OK? Can you say how many cells on day 2, 3 and 4 should be in a colony with a good grows or you may say that colonies with 30 cells at day 4 are stop growing ... 
% What are the parameters to mark the start of differentiation - the number of nuclei ? siuzes?

% Although our current analysis was performed on still images of colonies, this does not restrict their applicability for the analysis colonies under induced differentiation.

% \section*{Conclusion}

% We have quantified the structural properties of human embryonic stem cell colonies growing in Matrigel, at different days of plating. We applied tools used in image analysis and granular materials to demonstrate that the cell population evolved towards a segregated population in terms of the nucleus size. At the colony level, our analysis indicates a change in the nucleus sizes as the colonies grow in size. 

% We detected that small colonies formed at later stages (day 3 and 4 after plating) contained cells with larger nucleus still showing some features of pluripotent stem cells, see table 
% \ref{table1}. 

\section*{Data Availability}

The datasets used and/or analysed during the current study are available from the corresponding author on reasonable request.

\section*{Author Contributions}

SOF, IN, NGP, RAB, ML and AS designed the research. SOF designed and developed the computational framework and analysed the data. IN conceived, planned and performed the experiments. SOF, IN and LEW wrote the manuscript. All authors contributed critically to the drafts and gave final approval for publication.

\section*{Acknowledgments}
SOF acknowdledge financial support from the Consejo Nacional de Ciencia y Tecnología (CONACyT, Mexico) for the grant CVU-174695. ML acknowledges BBSRC UK (BB/I020209/1) and the H02020 ERC (614620) fellowships for providing financial support for this work. RAB wants to acknowledge financial support from CONACyT (Mexico) through the project 283279.

% \bibliographystyle{apalike}
% \bibliography{bibColonyFor}
\bibliographystyle{model2-names.bst}\biboptions{authoryear}
% \bibliography{bibColonyFor.bib}

\newpage
% % Uncomment if using bibtex (default)
% %\bibliographystyle{unsrt}      % mathematics and physical sciences
% %\bibliography{bibliographyBJ.bib}
% %\bibliographystyle{model2-names.bst}\biboptions{authoryear}

% \bibliography{bibliographyBJ.bib}
% \printbibliography
% % Uncomment if using biblatex
%  %\printbibliography
\newpage
%\onecolumn
\section{Supplementary Material}
\label{SuppMaterial}

% An online supplement to this article can be found by visiting BJ Online at \url{http://www.biophysj.org}.

\subsection{Nuclei detection}

\begin{figure*}
\begin{center}
    \includegraphics[width=0.95\textwidth]{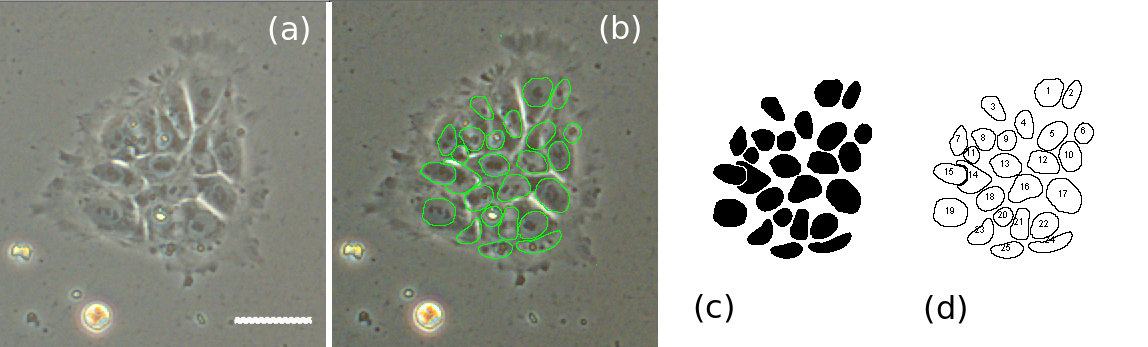}
\caption{(a) Example of the cell nuclei detection in a hESC colony. (b) The tracing was performed by outlining the nuclei. (c) This information was extracted and transformed in a binary file. (d) Using the plugin "Analyze Particles" in ImageJ \citep{Imagej} we detected each particle and measured the properties mentioned in the text. Scale bar $50 \, \mu$m.}
\label{manualTrace}
\end{center}
\end{figure*}

The analysis of each colony was performed using ImageJ \citep{Imagej} and the statistical analysis performed with Matlab\textregistered 2016b. Each cell was manually traced and processed using ImageJ as it is shown in Figure \ref{manualTrace}. We obtained the nuclei area $\alpha$ and perimeter $p$, alongside the properties that we describe in the following.

 \subsubsection*{Aspect Ratio}

The aspect ratio $\eta_\text{AR}$ is the relationship between the height and width of the nucleus. For a perfect circle, we obtain $\eta_\text{AR}=1$, elongated particles have $\eta_\text{AR}>1$. It is measured following the equation, $\eta_\text{AR} = \frac{a}{b}$, where $a$ is the major axis and $b$ the  minor axis of a rectangle that encloses the nucleus, in terms of the minimum area of enclosure. 

The Figure \ref{ShapeDescriptorsNuclei}(a) shows the PDF for the nuclei aspect ratio $\eta_\text{AR}$ following the colonies sampling day. Overall, the majority of cells have an aspect ratio of $\eta_\text{AR}$<2. Cells with elongated shapes $\eta_\text{AR} > 3$ are very few. This supports the applicability of the VD for an accurate tessellation of the colony.

 \subsubsection*{Perimeter}

The probability density function for nuclei perimeter is shown in Figure \ref{ShapeDescriptorsNuclei}(b). It has a mean of $\sim 50 \, \mu$m, and this value is similar for all sampling days.

\subsubsection*{Feret's Diameter }

The Feret's diameter ($\eta_\text{Feret}$), also know as maximum caliper measures the longest distance between any two points along the nucleus boundary. The probability density function (PDF)  for the Feret's diameter for the nuclei are shown in Figure \ref{ShapeDescriptorsNuclei}(c), according to the sampling day. The PDFs have a mean  $\sim 21 \, \mu$m, with the maximum caliper being $\sim  50 \, \mu$m.

\subsubsection*{Circularity}

Circularity is a shape descriptor that indicates the degree of similarity with a circle, therefore as this quantity approaches 0, the shape is less circular. It is calculated using the equation $\eta_\phi = \frac{4 \pi \alpha}{p^2}$, with $\alpha$ and $p$ the nucleus area and perimeter respectively. The values for the PDF of the circularity are shown in Figure \ref{ShapeDescriptorsNuclei}(d), with all three distributions centred around $\sim 0.85$, therefore the nuclei shapes are highly circular.

 \subsubsection*{Roundness}

It is very similar to circularity but is insensitive to irregular borders along the perimeter. It is measured using the highest axis of the best fit ellipse. The equation for the roundness is $\eta_\mathcal{R} = \frac{4 \alpha}{\pi a^2}$. The results for the PDF for the roundness are shown in Figure \ref{ShapeDescriptorsNuclei}(e).

 \subsubsection*{Solidity}

\begin{figure*}
\begin{center}
    \includegraphics[width=0.9\textwidth]{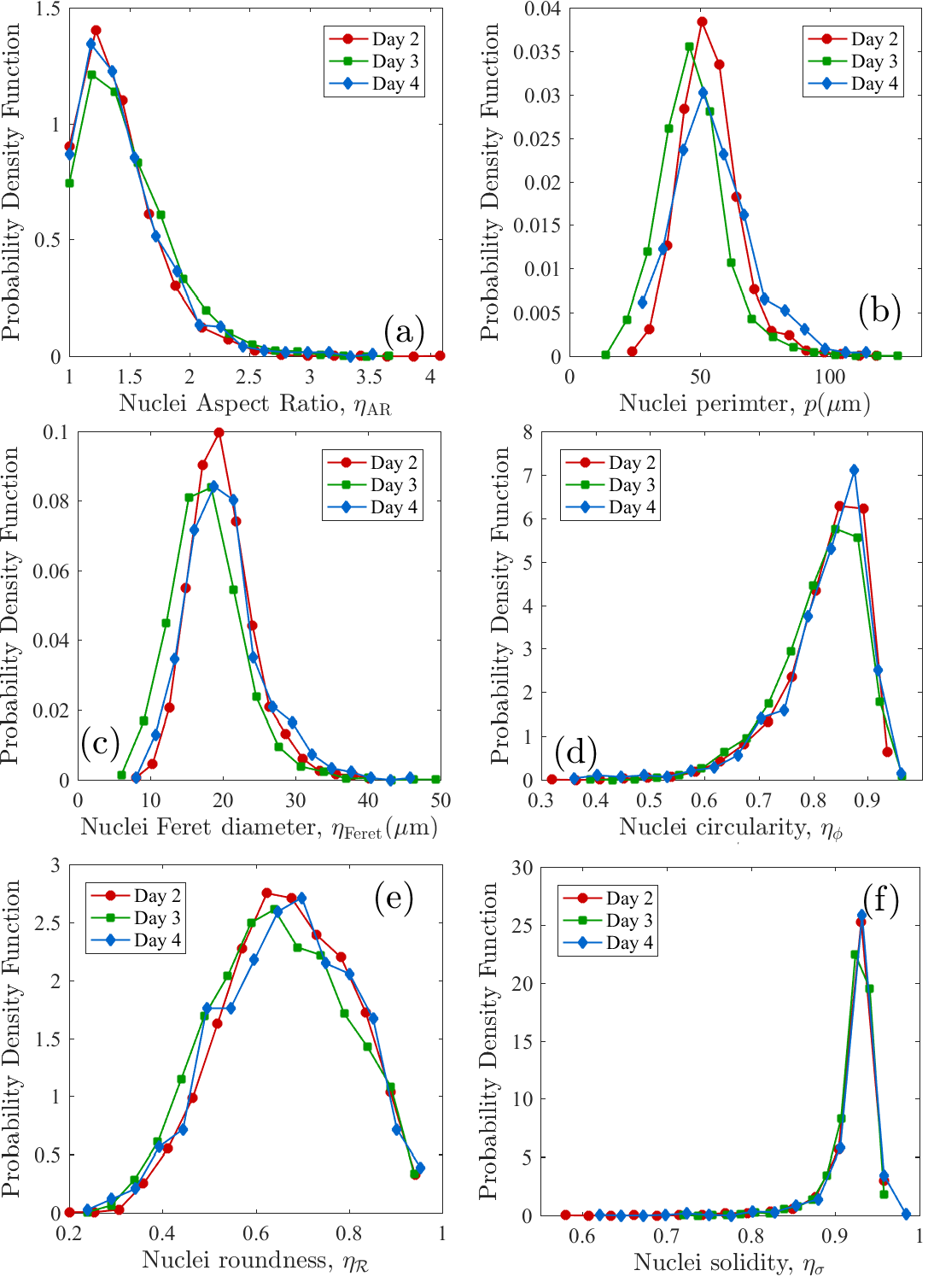}
\end{center}
\caption{Probability density functions (PDF) for the nuclei shape descriptors: (a) aspect ratio $\eta_\text{AR}$, (b) perimeter $p$, (c) Feret's diameter $\eta_\text{Feret}$, (d) circularity $\eta_\phi$, (e) roundness $\eta_\mathcal{R}$ and (f) solidity $\eta_\sigma$.}
\label{ShapeDescriptorsNuclei}
\end{figure*}

Describes the extent to which a shape is concave or convex. The solidity of a completely convex shape is 1, the farther the solidity deviates from 1, the concavity in the nucleus increases. It is calculated following the equation, $\eta_\sigma = \frac{\alpha}{\mathcal{A}}$, with $\mathcal{A}$ representing the area of the convex hull that best encloses the nucleus boundary. The results are shown in Figure \ref{ShapeDescriptorsNuclei}(f) and indicate that most of the nuclei have values close to 1.

\begin{figure*}
  \begin{center}
    \includegraphics[width=0.8\textwidth]{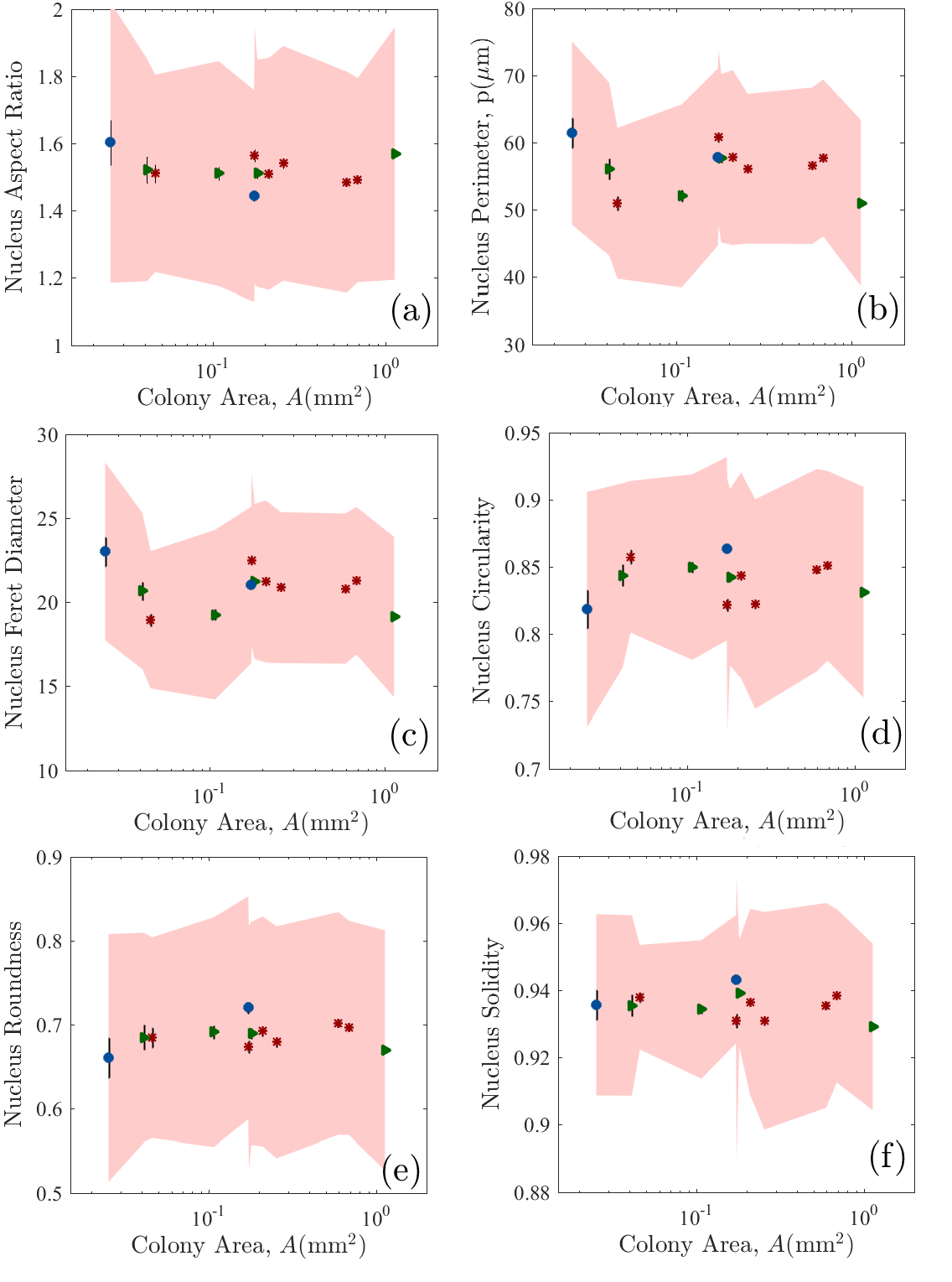}
\caption{Mean values of the shape descriptors obtained for the nuclei of the cells. The results are shown according to the sampling day, see the legend in f. The average value for the nucleus aspect ratio is $\eta_\text{AR} \sim 1.5$. The nuclei perimeter $p$ is between the range $[50, 60]$. The Feret's diameter $\eta_\text{Feret}$ is highest for the colonies with $A \approx 0.025 \ \text{mm}^2$, that contain differentiated cells at day 4, (blue {\color{myblue} $\bullet$} to the left). The perimeter $p$ decreases for larger colonies. A similar trend is shown in c for the Feret's diameter. The remaining shape descriptors corroborate that, in general, most of the nuclei are round and circular in shape, without irregular borders. }
\label{ShapeDesc}      
    \end{center}
\end{figure*}

The averages for each shape descriptor are shown in Table \ref{shapedescriptors} and the plots in Figure \ref{ShapeDesc}. The standard deviations range is the shaded region. The standard error of the mean is shown as a black line around the data point. Data points are presented with different symbols according to the day at which the image was taken (see legend panel f).

\begin{table*}
 
 \caption{Measurements obtained for the nuclei morphology and cellular parametric characteristics in hESC colonies.  The total number of cells in the colonies $N_c$, the mean cell nucleus area $\langle \alpha \rangle$, the mean number of nearest neighbours $\langle N_n \rangle$ and the mean intracellular distance $\langle \ell_n \rangle$, alongside the standard deviations for the measurements.}
 
 \begin{center}
     \begin{tabular}{| c | c | c | c | c | c |}
      \hline
 $A$ ($\text{mm}^2$) & $N_c$ & $\langle \alpha \rangle, \ \mu \mbox{m}^2$ &  $\langle N_n \rangle$  & $\langle \ell_n \rangle, \ \mu \mbox{m}$  & Day \\ \hline \hline
0.025 $\pm$ 0.007$^\dag$  &	39  $\pm$ 10   & 269  $\pm$ 111 & 4.81$\pm$ 1.40 &	22.16 $\pm$ 3.30 & 4 \\ \hline
0.041 $\pm$ 0.005         &	71  $\pm$ 34   & 212  $\pm$ 105  & 5.08 $\pm$ 1.41 &	20.48 $\pm$ 3.11 & 3 \\ \hline
0.046 $\pm$ 0.002         &	115	$\pm$ 1.4  & 185  $\pm$ 82   & 5.20 $\pm$ 1.44 &	20.31 $\pm$ 3.73 & 2 \\ \hline
0.107  &	305	          & 206	$\pm$ 96   & 5.57 $\pm$ 1.23 &	18.43 $\pm$ 2.61 & 3 \\ \hline
0.173  &	363	          & 239	$\pm$ 109  & 5.61 $\pm$ 1.15 &	21.60 $\pm$ 3.13 & 4 \\ \hline
0.174  &	375           &	248	$\pm$ 100  & 5.63 $\pm$ 1.16 &	21.46 $\pm$ 3.10 & 2 \\ \hline
0.180  &	409	          & 230 $\pm$ 104  & 5.60 $\pm$ 1.26 &	21.29 $\pm$ 3.06 & 3 \\ \hline
0.210  &	514	          & 232	$\pm$ 99   & 5.67 $\pm$ 1.14 &	20.85 $\pm$ 3.06 & 2 \\ \hline
0.255  &	543	          & 226	$\pm$ 80   & 5.62 $\pm$ 1.28 &	20.67 $\pm$ 3.14 & 2 \\ \hline
0.596  &	1026          &	222	$\pm$ 92   & 5.76 $\pm$ 1.12 &	20.37 $\pm$ 3.15 & 2 \\ \hline
0.690  &	1489          &	211	$\pm$ 91   & 5.84 $\pm$ 1.03 &	20.13 $\pm$ 3.08 & 2 \\ \hline
$1.131^\ddag$      &1982  & 179	$\pm$ 87   & 5.91 $\pm$ 0.96 &	18.56 $\pm$ 3.11 & 3 \\ 
 \hline
\end{tabular}

\label{tableres}
\end{center}

% \begin{tablenotes}
% \item[a]$\dag$  For the smallest colonies we obtained statistics from several samples, grouping them according to imaging day.
% \item[b]$\ddag$  Only a portion of the colony was analysed.
% \end{tablenotes}

\end{table*}

\begin{table*}

 \caption{Shape descriptors obtained for the nuclei features in hESC colonies: aspect ratio $\eta_\text{AR}$, perimeter $p$, Feret's diameter $\eta_\text{Feret}$, roundness $\eta_\mathcal{R}$ and solidity $\eta_\sigma$.}
 
 \begin{center}
     \begin{tabular}{| c | c | c | c | c | c | c | c |}
      \hline
 $A$ ($\text{mm}^2$) &  $\eta_\text{AR}$ &  $p$  & $\eta_\text{Feret}$ & $\eta_\phi$ & $\eta_\mathcal{R}$ & $\eta_\sigma$ & Day \\ \hline
0.025 $\pm$ 0.007$^\dag$  & 1.60 $\pm$ 0.42 &	61.41 $\pm$	13.66 &	23.02 $\pm$	5.30 &	0.82 $\pm$	0.09 &	0.66 $\pm$	0.15 &	0.95 $\pm$	0.03 & 4 \\ \hline
0.041 $\pm$ 0.005 & 1.52 $\pm$	0.33 &	56.06 $\pm$	12.89 &	20.65 $\pm$	4.66 &	0.84 $\pm$	0.07 &	0.69 $\pm$	0.12 &	0.93 $\pm$	0.03 & 3 \\ \hline
0.046 $\pm$ 0.002 & 1.51 $\pm$	0.29 &  50.96 $\pm$	11.23 &	18.94 $\pm$	4.09 &  0.86 $\pm$	0.06 &	0.68 $\pm$	0.12 &	0.94 $\pm$	0.02 & 2 \\ \hline
0.107                & 1.51 $\pm$	0.33 &	52.06 $\pm$ 13.64 & 19.25 $\pm$	5.06 &  0.85 $\pm$	0.07 &	0.69 $\pm$	0.14 &	0.93 $\pm$	0.02 & 3 \\ \hline
0.173                & 1.44 $\pm$	0.31 &	57.85 $\pm$	13.19 &	21.02 $\pm$	4.67 &	0.86 $\pm$	0.07 &	0.72 $\pm$	0.13 &	0.94 $\pm$	0.02 & 4 \\ \hline
0.174                & 1.56	$\pm$   0.38 &	60.75 $\pm$	13.05 &	22.50 $\pm$	5.14 &	0.82 $\pm$	0.09 &	0.67 $\pm$	0.14 &	0.93 $\pm$	0.04 & 2 \\ \hline
0.180                & 1.51 $\pm$	0.34 &	57.66 $\pm$	12.56 &	21.23 $\pm$	4.62 &	0.84 $\pm$	0.07 &	0.69 $\pm$	0.13 &	0.94 $\pm$	0.01 & 3 \\ \hline
0.210                & 1.51	$\pm$   0.35 &	57.78 $\pm$	13.03 &	21.22 $\pm$	4.83 &	0.84 $\pm$	0.08 & 	0.69 $\pm$	0.14 &	0.94 $\pm$	0.03 & 2 \\ \hline
0.255                & 1.54 $\pm$  0.35 &	56.11 $\pm$	11.16 &	20.86 $\pm$ 4.50 &	0.82 $\pm$	0.08 &	0.68 $\pm$	0.14 &	0.93 $\pm$	0.03 & 2 \\ \hline
0.596                & 1.48 $\pm$  0.33	&   56.54 $\pm$ 11.65 & 20.80 $\pm$ 4.48 &  0.85 $\pm$	0.07 &	0.70 $\pm$	0.13 &	0.94 $\pm$	0.03 & 2 \\ \hline
0.690                & 1.49	$\pm$  0.30 &	57.70 $\pm$	11.69 &	21.27 $\pm$	4.40 &	0.85 $\pm$	0.07 &	0.70 $\pm$	0.13 &	0.94 $\pm$ 0.03 & 2 \\ \hline
$1.131^\ddag$        & 1.57 $\pm$  0.38 &	51.00 $\pm$	12.37 &	19.11 $\pm$	4.77  &	0.83 $\pm$	0.08 &	0.67 $\pm$	0.14 &	0.93 $\pm$	0.02 & 3 \\ 
 \hline
\end{tabular}

\label{shapedescriptors}
\end{center}

\begin{tablenotes}
\item[a]$\dag$  For the smallest colonies we obtained statistics from several samples, grouping them according to imaging day.
\item[b]$\ddag$  Only a portion of the colony was analysed.
\end{tablenotes}

\end{table*}

The behaviour of the shape descriptor, although useful at the colony level for the detection of morphological changes between pluripotent and differentiated colonies, do not show a significant change at the nucleus level. The aspect ratio of the nuclei is $\sim 1.5$ in all cases, see Figure \ref{ShapeDesc}(a).

\subsection{Edge detection}

\begin{figure}
\begin{center}
    \includegraphics[width=0.4\textwidth]{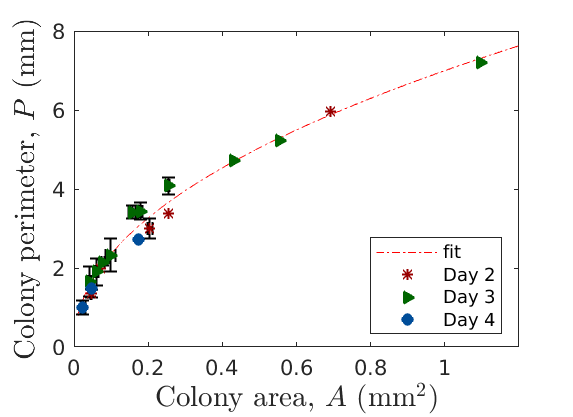}
\end{center}
\caption{Colony perimeter $P$ as a function of the colony area $A$. These datapoints were obtained by applying the canny Deriche algorithm to the samples. The red dashed-dotted line shows the best fit to a power function with a scaling factor $\kappa=7.207$ and exponent $\gamma=0.47$, ($\text{R}^2$ = 0.963).}
\label{Deriche}
\end{figure}

To detect the colony edges from the phase contrast images we use the canny Deriche algorithm included in ImageJ \citep{Imagej} to aid in the identification of the border as shown in Figure \ref{Deriche}(a) and (b). To obtain the boundary we traced manually the borders obtained in (b).

\begin{figure*}
\begin{center}
    \includegraphics[width=0.95\textwidth]{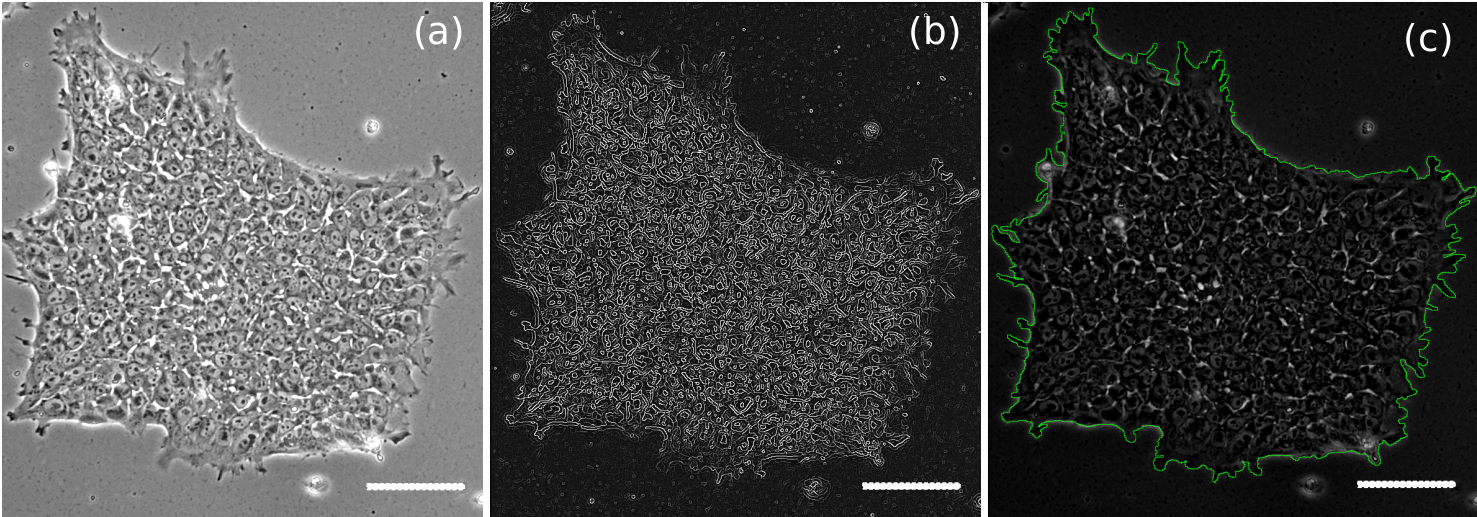}
\end{center}
\caption{(a) Human ESCs colony at day 3 processed with (b) the canny Deriche algorithm to obtain (c) the boundary.}
\label{DericheDet}
\end{figure*}

% \subsubsection*{Colonies Aspect Ratio}

\begin{figure*}
\begin{center}
    \includegraphics[width=0.95\textwidth]{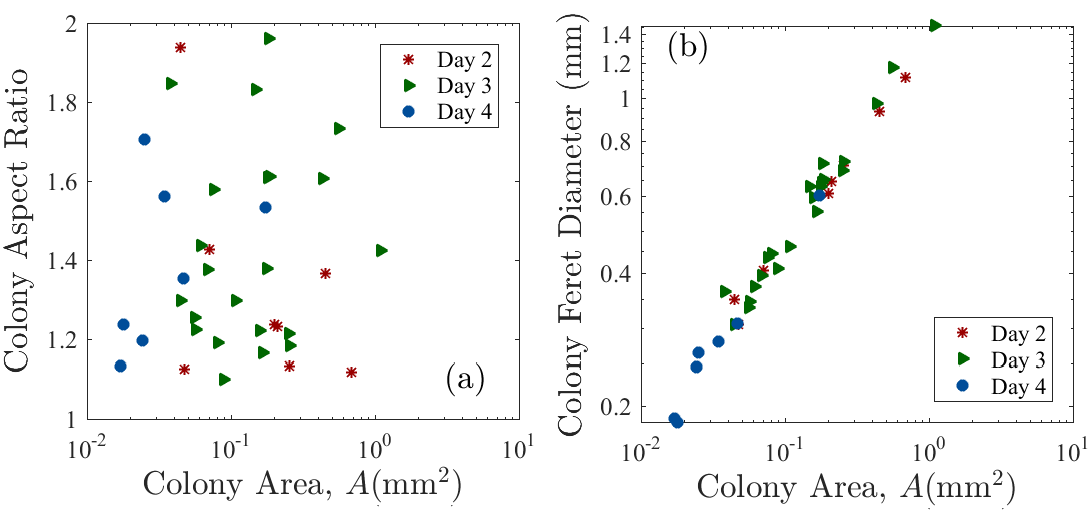} 
\end{center}
\caption{Numerical values of the (a) aspect ratio and (b) the Feret's diameter for the colonies shown in the Table \ref{tableres2}. }
\label{AspectRatioCol}
\end{figure*}

\subsection{Colony reconstruction}

\begin{figure*}
\begin{center}
    \includegraphics[width=0.4\textwidth]{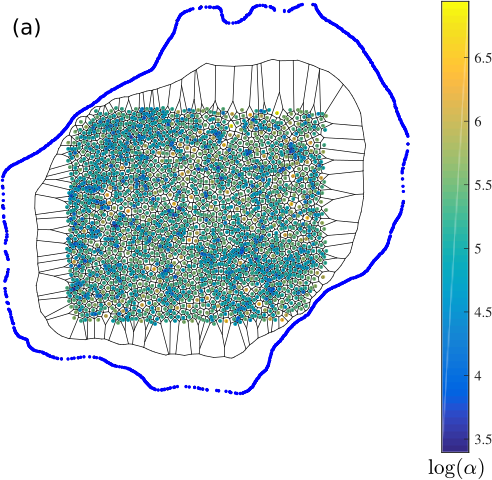}     \includegraphics[width=0.45\textwidth]{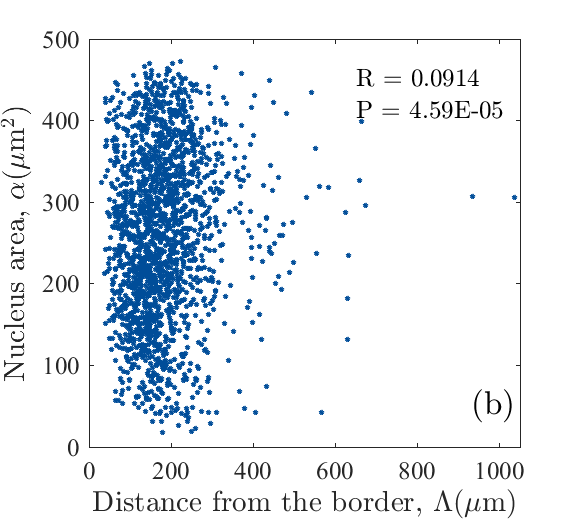}
\end{center}
\caption{(a) Voronoi tessellation constructed for the colony shown in Figure \ref{scheme_outline}, with an area $A = 1.131 \, \text{mm}^2$ and  $N_c = 1982$. The border of the colony is shown with a blue dotted line. (b) Cell nucleus area ($\alpha$) as a function of their distance $\Lambda \, (\mu$m) from the colony edge. The correlation coefficient is $R = 0.091$ and indicates a null dependence between both variables.}
\label{DistBorder}
\end{figure*}

To measure the correlation between the cell nucleus size and its position within the colony we reconstructed the colony shown in Figure \ref{scheme_outline} using the Voronoi diagram (VD) bounded by the colony's boundaries. In Figure \ref{DistBorder}(a) we show the reconstructed VD for the colony in Figure \ref{scheme_outline}. Using the centroid position for each cell we calculated their (closest) distance $\Lambda$ from the boundary and plotted this as a function of the cell nucleus area $\alpha$, see Figure \ref{DistBorder}(b). Our results indicate that the cell sizes are not correlated to their position within the colony, with a correlation coefficient of $R = 0.091$.

 % \begin{figure}
 %     \includegraphics[width=0.45\textwidth]{exceedingly.png}
 % \caption{Human embryonic stem cell colonies at day 3 after plating. Some cells are showing a large nucleus while keeping a round shape with well-defined nucleoli.}
 % \label{largeExcee}      
 % \end{figure}

\subsubsection*{Structural analysis through the radial distribution function}

Radial distribution functions (RDF) are tools used widely in crystallography and soft condensed matter physics to characterise the structure of different materials \citep{SST2004}. The RDF, denoted with $g(r)$, determines the number of particles in a spherical shell of radius $r$ and thickness $dr$, i.e., $n g(r) 4 \pi r^2 dr$. In other words, it describes the variation of the local cell density within a distance $r$ as viewed from the centring cell, relative to its bulk value. For ordered materials, such as crystals, the radial distribution function show an oscillating behaviour, where the peaks in $g(r)$ are interpreted as the average inter-particle distances. The information contained in the RDF is a spatial average and has its limitations when the system is not isotropic.

 % \begin{minipage}{\linewidth}
 %      \centering
 %      \begin{minipage}{0.45\linewidth}
 %          \begin{figure*}[H]
 %               \includegraphics[width=0.4\textwidth]{CorrelationgR.png}
	% \caption{Radial distribution functions $g(r)$ for the two largest colonies analysed. The first peak, associated with the separation of the first nearest neighbours indicates a location of < $20 \, \mu$m, consistent with the results presented in Figure \ref{plotsTable} obtained through the Voronoi diagram.}
	% \label{gradial}
 %          \end{figure*}
 %      \end{minipage}
 %      \hspace{0.05\linewidth}
 %      \begin{minipage}{0.45\linewidth}
 %          \begin{figure*}[H]
 %              \includegraphics[width=0.4\textwidth]{AreavsPerim.png}
	% \caption{Colony perimeter $P$ as a function of the colony area $A$. These datapoints were obtained by applying the canny Deriche algorithm to the samples. The red dashed-dotted line shows the best fit to a power function with a scaling factor $\kappa=7.207$ and exponent $\gamma=0.47$, ($\text{R}^2$ = 0.963).}
	% \label{Deriche}
 %          \end{figure*}
 %      \end{minipage}
 %  \end{minipage}

\begin{figure}
  \begin{center}
    \includegraphics[width=0.4\textwidth]{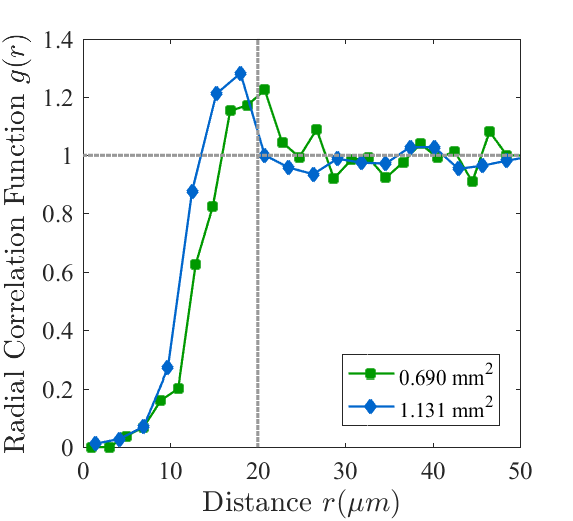}
\caption{Radial distribution functions $g(r)$ for the two largest colonies analysed. The first peak, associated with the separation of the first nearest neighbours indicates a location of < $20 \, \mu$m, consistent with the results presented in Figure \ref{plotsTable} obtained through the Voronoi diagram.}
\label{gradial}      
    \end{center}
\end{figure}

Figure \ref{gradial} show results for the RDF for two colonies of different sizes with areas $0.690 \, \text{mm}^2$ and $1.131 \, \text{mm}^2$, respectively. For both cases, the first peak is the best-defined one and corresponds to the distribution of the distance between the first nearest neighbours $\ell_1 \sim 18.56 \, \mu$m  and $\ell_1 \sim 18.02 \, \mu$m respectively. The position of the second peak, gives the average distance or coordination, between second neighbours $\ell_2$. We conclude that the colonies show a short-range order and the nearest coordination shells are visible in both cases. But, as we increase $r$ to account for the second nearest neighbours the second peak is washed out and broader than the first due to missing long-range order. The largest colony shows a second peak centred around $\ell_2 \sim 40.2 \, \mu$m. Interestingly, the RDF shows that the cells in the largest colony ($A = 1.131 \, \text{mm}^2$)  are more packed since the blue curve crosses the horizontal dotted line before the green curve.  The results are shown in Figure \ref{gradial} and indicate that $g(r)$ is similar to the ones obtained for amorphous materials, therefore the structure of the material becomes blurred as the radius $r$ increases.

In biology, calculations of $g(r)$ have been performed to study protein organisation, aggregation of particles on cell membranes, \citep{Pearson79,Muller84}.  This tool has not been used previously in the literature to characterise dense aggregates of cells.

% \subsection*{Datasets}

% In section \ref{NucleiMorph} we presented the statistical properties obtained for the cell's nucleus, such as their mean area $\langle \alpha \rangle$, mean number of nearest neighbours $\langle N_n \rangle$ and mean distance to nearest neighbours $\ell_n$. The 

% \begin{landscape}

\begin{table*}
 
 %Colony	Area	NC	Perim	Circ	Feret	MinFeret	AR	Round	Sol	DAY
 \caption{Datasets of the morphological and parametric characteristics for hESC colonies.  We show the colony area $A$, number of cells $N_c$, perimeter $\mbox{P}$, circularity $\Phi$, Feret diameter, minimum Feret diameter, aspect ratio $\mbox{L}$, roundness $\mathcal{R}$, solidity $\Sigma$ and day of imaging.}
 
 \begin{center}
     \begin{tabular}{| c | c | c | c | c | c | c | c | c |c | c |}
      \hline
Identifier & $A$($\text{mm}^2$) & $N_c$ & $\mbox{P}$ ($\mu \text{m}$) & $\Phi$ & Feret & MinFeret & $\mbox{L}$ & $\mathcal{R}$ &  $\Sigma$ & Day \\ \hline \hline
{\scriptsize DAY2\_4x5}    &	0.691  &	1489  &	5.960 &	0.245   &	1115.1  &	953.2 &	1.117 &	0.895 &	0.874 &	2 \\ \hline
{\scriptsize DAY2\_6x10}   &	0.456  &	1026  &	2.747 &	0.759   &	 931.6  &	691.4 &	1.365 &	0.733 &	0.962 &	2 \\ \hline
{\scriptsize DAY2\_7x10}   &	0.209  &	514   &	3.180 &	0.260   &	 644.8 &	513.1 &	1.233 &	0.811 &	0.840 &	2 \\ \hline
{\scriptsize DAY2\_8Ax10}  &	0.047  &	114   &	1.428 &	0.292   &	 305.6 &	252.3 &	1.125 &	0.889 &	0.810 &	2 \\ \hline
{\scriptsize DAY2\_8Bx10}  &	0.045  &	116   &	1.283 &	0.341   &	 348.5 &	190.7 &	1.938 &	0.516 &	0.845 &	2 \\ \hline
{\scriptsize DAY2\_9x10}   &	0.071  &	206   &	1.979 &	0.228   &	 405.0 &	275.3 &	1.428 &	0.700 &	0.827 &	2 \\ \hline
{\scriptsize DAY2\_10x10}  &	0.254  &	543   &	3.390 &	0.278   &	 702.4 &	573.3 &	1.132 &	0.883 &	0.883 &	2 \\ \hline
{\scriptsize DAY2\_11x10}  &	0.199  &	375   &	2.825 &	0.314   &	 607.3 &	481.8 &	1.238 &	0.808 &	0.872 &	2 \\ \hline
{\scriptsize DAY3\_1x10}   &	0.164  &	457   &	3.249 &	0.196   &	 553.2 &	452.5 &	1.165 &	0.858 &	0.842 &	3 \\ \hline
{\scriptsize DAY3\_2X10}   &	0.056  &	151   &	1.770 &	0.226   &	 344.5 &	279.6 &	1.224 &	0.817 &	0.816 &	3 \\ \hline
{\scriptsize DAY3\_3x10}   &	0.255  &	862   &	3.930 &	0.208   &	 716.1 &	555.3 &	1.183 &	0.846 &	0.845 &	3 \\ \hline
{\scriptsize DAY3\_4x10}   &	0.107  &	305   &	2.616 &	0.197   &	 460.5 &	356.8 &	1.298 &	0.770 &	0.873 &	3 \\ \hline
{\scriptsize DAY3\_5x10}   &	0.080  &	198   &	2.187 &	0.211   &	 444.4 &	341.4 &	1.192 &	0.839 &	0.763 &	3 \\ \hline
{\scriptsize DAY3\_6x5}    &	1.097  &	-     &	7.205 &	0.263   &	1460.6 &   1040.9 &	1.420 &	0.703 &	0.888 &	3 \\ \hline
{\scriptsize DAY3\_7x10}   &	0.252  &	703   &	4.230 &	0.177   &	 685.4 &	529.9 &	1.213 &	0.824 &	0.848 &	3 \\ \hline
{\scriptsize DAY3\_10x10A} &	0.056  &	83    &	1.535 &	0.299   &	 334.2 &	275.7 &	1.255 &	0.797 &	0.839 &	3 \\ \hline
{\scriptsize DAY3\_10x10B} &	0.038  &	74    &	1.922 &	0.129   &	 363.6 &	215.7 &	1.848 &	0.541 &	0.651 &	3 \\ \hline
{\scriptsize DAY3\_11x10}  &	0.088  &	229   &	2.023 &	0.270   &	 410.0 &	341.6 &	1.100 &	0.909 &	0.837 &	3 \\ \hline
{\scriptsize DAY3\_12x10}  &	0.069  &	166   &	1.983 &	0.219   &	 394.8 &	289.6 &	1.376 &	0.727 &	0.803 &	3 \\ \hline
{\scriptsize DAY3\_14x10}  &	0.182  &	440   &	3.664 &	0.171   &	 710.0 &	391.8 &	1.960 &	0.510 &	0.844 &	3 \\ \hline
{\scriptsize DAY3\_16x10}  &	0.159  &	359   &	3.410 &	0.172   &	 594.0 &	462.6 &	1.223 &	0.818 &	0.778 &	3 \\ \hline
{\scriptsize DAY3\_17x5}   &	0.430  &	-     &	4.718 &	0.243   &	 974.8 &	614.9 &	1.607 &	0.622 &	0.876 &	3 \\ \hline
{\scriptsize DAY3\_18x10}  &	0.175  &	-     &	3.320 &	0.200   &	 629.9 &	461.7 &	1.379 &	0.725 &	0.830 &	3 \\ \hline
{\scriptsize DAY3\_19x10}  &	0.076  &	-     &	2.095 &	0.217   &	 433.9 &	292.2 &	1.579 &	0.633 &	0.775 &	3 \\ \hline
{\scriptsize DAY3\_20x10}  &	0.180  &	409   &	3.222 &	0.218   &	 643.6 &	439.3 &	1.609 &	0.622 &	0.833 &	3 \\ \hline
{\scriptsize DAY3\_21x10}  &	0.147  &	-     &	3.593 &	0.143   &	 629.5 &	380.3 &	1.831 &	0.546 &	0.792 &	3 \\ \hline
{\scriptsize DAY3\_23x5}   &	0.555  &	-     &	5.232 &	0.255   &	1172.1 &	746.7 &	1.734 &	0.577 &	0.857 &	3 \\ \hline
{\scriptsize DAY3\_24x10}  &	0.061  &	-     &	2.341 &	0.140   &	 372.2 &	291.9 &	1.438 &	0.695 &	0.764 &	3 \\ \hline
{\scriptsize DAY3\_25x10}  &	0.045  &	69    &	1.390 &	0.291   &	 305.2 &	231.7 &	1.297 &	0.771 &	0.858 &	3 \\ \hline
{\scriptsize DAY3\_26x10}  &	0.184  &	-     &	3.560 &	0.183   &	 654.2 &	434.8 &	1.610 &	0.621 &	0.839 &	3 \\ \hline
{\scriptsize DAY4\_18x10}  &	0.017  &	25    &	0.871 &	0.281   &	 187.7 &	157.7 &	1.132 &	0.883 &	0.825 &	4 \\ \hline
{\scriptsize DAY4\_3x10}   &	0.173  &	363   &	2.734 &	0.290   &	 600.6 &	413.2 &	1.534 &	0.652 &	0.862 &	4 \\ \hline
{\scriptsize DAY4\_7x10}   &	0.047  &	86    &	1.474 &	0.270   &	 307.1 &	224.6 &	1.354 &	0.739 &	0.844 &	4 \\ \hline
{\scriptsize DAY4\_8x10}   & 	0.024  &	40    &	1.237 &	0.200   &	 245.1 &	173.2 &	1.197 &	0.836 &	0.766 &	4 \\ \hline
{\scriptsize DAY4\_9x10}   &	0.035  &	46    &	1.083 &	0.372   &	 280.2 &	188.7 &	1.562 &	0.640 &	0.869 &	4 \\ \hline
{\scriptsize DAY4\_11x10}  &	0.025  &	46    &	1.064 &	0.279   &    263.2 &	162.8 &	1.706 &	0.586 &	0.765 &	4 \\ \hline
{\scriptsize DAY4\_21x10}  &	0.018  &	52    &	0.791 &	0.359   &	 184.1 &	157.7 &	1.236 &	0.809 &	0.827 &	4 \\ 
\hline
\end{tabular}

\label{tableres2}
\end{center}

\begin{tablenotes}
\item[a]$-$  Cells were not counted, only features analysed.
% \item[b]$\ddag$  Only a portion of the colony was analysed.
\end{tablenotes}

\end{table*}
% \end{landscape}

% \begin{figure}
%     \begin{center}
%         \includegraphics[width=0.9\textwidth]{sameAreatags.png}
%     \end{center}
%     \caption{Colonies with areas (a) $A = 0.174$, (b) 0.107$ and (c) $0.173$ (with $N_c = 375, \, 305$ and $363$ cells) analysed at days 2, 3 and 4, respectively. Since (a) and (c) have approximately the same area  }
% \label{sameAreas}      
% \end{figure}

\end{document}